\documentclass{aastex61}

\usepackage{natbib}
\usepackage{array}\usepackage{makecell}
\usepackage{amsmath}
\usepackage{graphicx}
\definecolor{orange}{rgb}{.8,0.4,0}

\shorttitle{Transmission Spectra of Kepler 51b and 51d}
\shortauthors{Libby-Roberts et al.}

\newcommand{\kepler}{{\em Kepler }}

\newcommand{\HST}{{\em HST }}

\begin{document}

\title{The Featureless Transmission Spectra of Two Super-Puff Planets}

\correspondingauthor{Jessica E. Libby-Roberts}
\email{jessica.e.roberts@colorado.edu}


\author[0000-0002-2990-7613]{Jessica E. Libby-Roberts}
\affil{Department of Astrophysical and Planetary Sciences, University of Colorado, Boulder, CO 80309, USA}

\author[0000-0002-3321-4924]{Zachory K. Berta-Thompson}
\affil{Department of Astrophysical and Planetary Sciences, University of Colorado, Boulder, CO 80309, USA}

\author[0000-0002-0875-8401]{Jean-Michel D\' esert}
\affil{Anton Pannekoek Institute for Astronomy, University of Amsterdam, 1090 GE Amsterdam, Netherlands}

\author[0000-0003-1298-9699]{Kento Masuda}
\affil{Department of Astrophysical Sciences, Princeton University, 4 Ivy Lane, Princeton, NJ, 08544, USA}
\affil{NASA Sagan Fellow}

\author[0000-0002-4404-0456]{Caroline V. Morley}
\affil{The University of Texas at Austin, Austin, TX, USA}

\author{Eric D. Lopez}
\affil{NASA Goddard Space Flight Center, 8800 Greenbelt Rd, Greenbelt, MD 20771, USA}
\affil{GSFC Sellers Exoplanet Environments Collaboration, NASA GSFC, Greenbelt, MD 20771}

\author{Katherine M. Deck}
\affil{Division of Geological and Planetary Sciences, California Institute of Technology, Pasadena, CA 91101, USA}

\author[0000-0003-3750-0183]{Daniel Fabrycky}
\affil{Department of Astronomy and Astrophysics, University of Chicago, Chicago, IL 60637, USA}

\author[0000-0002-9843-4354]{Jonathan J. Fortney}
\affil{Department of Astronomy and Astrophysics, University of California, Santa Cruz, CA 95064, USA}

\author[0000-0002-2338-476X]{Michael R. Line}
\affil{School of Earth and Space Exploration, Arizona State University, Tempe, AZ 85281, USA}

\author{Roberto Sanchis-Ojeda}
\affil{Netflix, 100 Winchester Circle, Los Gatos, CA 95032 USA}

\author[0000-0002-4265-047X]{Joshua N.\ Winn}
\affil{Department of Astrophysical Sciences, Princeton University, 4 Ivy Lane, Princeton, NJ 08544 USA}




\begin{abstract}

The {\it Kepler} mission revealed a class of planets known as ``super-puffs,'' with masses only a few times larger than Earth's but radii larger than Neptune, giving them very low mean densities. All three of the known planets orbiting the young solar-type star Kepler 51 are super-puffs.
The Kepler 51 system thereby provides an opportunity for a comparative study of the structures and atmospheres of this
mysterious class of planets, which may provide clues about their formation and evolution. We observed two transits each of Kepler 51b and 51d with the Wide Field Camera 3 (WFC3) on the {\it Hubble Space Telescope}.
Combining new WFC3 transit times with re-analyzed {\it Kepler} data and updated stellar parameters, we confirmed that all three planets have densities lower than 0.1~g/cm$^{3}$.
We measured the WFC3 transmission spectra to be featureless between 1.15 and 1.63~$\mu$m, ruling
out any variations greater than 0.6 scale heights (assuming a H/He dominated atmosphere), thus showing no significant water absorption features.
We interpreted the flat spectra as the result of a high-altitude aerosol layer (pressure $<$3~mbar) on each planet.
Adding this new result to the collection of flat spectra that have been observed for other sub-Neptune planets,
we find support for one of the two hypotheses introduced by \citet{crossfield.and.kreidberg.2017}, that
planets with cooler equilibrium temperatures have more high-altitude aerosols. We strongly disfavor their other hypothesis that
the H/He mass fraction drives the appearance of large amplitude transmission features.

\end{abstract}




\section{Introduction} \label{sec:intro}

Kepler 51 is a moderately young (500 Myr) G-type star that hosts three Jupiter-sized planets with orbital periods of 45, 85, and 130 days \citep{steffen.et.al.2013}. The star is relatively faint \citep[$J=13.56$, $m_{\rm Kep}=14.669$;][] {cutri.et.al.2003,latham.et.al.2005}, making it difficult to measure the planet masses
with the Doppler technique using currently available instruments. However, with their near-resonant periods, Kepler 51b, 51c, and 51d display high S/N transit timing variations (TTVs) of 5--45 minutes \citep{masuda.2014}. From \kepler light curves, \citet{masuda.2014} determined all three planets had unusually low masses (less than $10~M_{\oplus}$) given their large radii of 6 to 10 $R_{\oplus}$. With densities less than 0.1~g/cm$^{3}$, the Kepler 51 planets are the lowest-density planets to date \added{according to the NASA Exoplanet Archive} \citep{archive.paper}. Not only does the Kepler 51 system present us with an opportunity to study unusually low-density planets, it also allows us to glimpse a snapshot into the evolution of ``teenage'' \added{($<$1 Gyr)} sub-Neptune mass planets.

With their extremely low densities, the Kepler 51 planets join the ranks of a rare class of exoplanets known as super-puffs \citep{lee.and.chiang.2016}. Super-puffs, \added{including Kepler 47c \citep{orosz.et.al.2019}, Kepler 79d \citep{jontof-hutter.et.al.2014}, and Kepler 87c \citep{ofir.et.al.2014},} are cooler and less massive than the inflated low-density hot-Jupiters. The thermal evolution models of \added{\citet{rogers.et.al.2011},} \citet{lopez.et.al.2012}, and  \citet{batygin.and.stevenson.2013} can reproduce such low-density exoplanets with large mass fractions of hydrogen and helium, but the process of forming low-mass, H/He-rich planets continues to present an interesting challenge. \citet{lee.and.chiang.2016} suggested that super-puffs
form beyond 1~AU in a \replaced{relatively cool disk with a low opacity}{region of the disk with a low opacity}, allowing the accreting envelopes of the planets to cool rapidly. This swift cooling allows the planets to accrete a large H/He atmosphere, despite having relatively
low-mass cores. Because the planets of Kepler 51 are near a 3:2:1 resonance, \citet{lee.and.chiang.2016} proposed that these planets migrated inwards to their current location. If this formation process is correct, then \citet{lee.and.chiang.2016} predict the atmospheres of the Kepler 51 planets should contain water delivered either by infalling icy planetesimals during formation or by \replaced{sublimation}{erosion} of their interior cores. \replaced{Observations of water features in the atmospheres of these planets would support their theory that these planets formed further out in the disk before undergoing migration.}{While observations of water features in the atmospheres of these planets would not be direct proof of this theory, it would support the hypothesis that these planets formed further out in the disk before undergoing migration.} \added{An alternate hypothesis presented by \citet{Millholland.2019} is that these planets do not have an unusually high H/He, but that their radii are instead inflated by obliquity tides. If this is the case, then the low-densities of Kepler 51 are a consequence of their continual interaction with the host star and not from their formation. However, for their current model, the Kepler 51 planets are too far away from the star so it is currently unknown as to the effectiveness of obliquity tides in this case.}

Kepler 51b and 51d are prime targets for searching for this water signal via transmission spectroscopy, despite the
faintness of the star. The low densities of these planets demand that their atmospheres be low mean molecular weight (H/He dominated, solar composition). Combining an assumed mean molecular weight of 2.3 amu, along with their equilibrium temperatures and their small surface gravities, we determined that Kepler 51b and 51d should possess scale heights between 2000 and 3000 km. Using the relationship that the change in transit depth due to a molecular feature scales as $2HR_{p}/R_{s}^{2}$ \citep{brown.2001}, we calculate that a one scale-height variation on either planet corresponds to a change of about 300~ppm in transit depth. Kepler 51c likely has a similar scale height, however the planet only grazes the star during transit making the variations in transit depth much smaller than Kepler 51b and 51d. In this paper, we use the {\it Hubble Space Telescope} (\HST) Wide Field Camera 3 (WFC3) to observe the transmission spectra of Kepler 51b and 51d, aiming to measure the amplitude of water absorption features.


Thus far, the masses of all super-puffs discovered were obtained via transit-timing variations (TTVs). Studies have suggested that TTV-determined masses, on average, tend to be smaller than average masses determined from radial velocity observations \citep[e.g.][]{steffen.2016, mills.and.mazeh.2017}. However, \citet{steffen.2016} concludes that this offset is likely influenced by the different selection biases between the radial velocity method (best for massive planets with short periods) and transit-timing method (best for large planets with long periods), not by incorrect mass estimates from any one method. \citet{mills.and.mazeh.2017} also compare the population of RV-determined mass planets to those planets with masses determined from TTV measurements and find no discrepancy between the two methods for comparable periods. \added{Both \citet{weiss.et.al.2017} and \citet{petigura.et.al.2017a} performed their own independent TTV and RV analysis on different systems and came to this same conclusion.}
Thus, the low TTV masses of super-puff planets are likely not a consequence of a flawed technique but rather \added{due} to the TTV method's ability to probe a different population of planets at longer periods and lower densities.

Here we present new {\it HST} transmission spectra of Kepler 51b and 51d, as well as a critical re-investigation of the low-densities for all three Kepler 51 planets. We outline our new Hubble/WFC3 observations in Section~\ref{sec:observation} and detail our analysis of both the white light curves and spectroscopic light curves in Section~\ref{sec:analysis}. \added{This section also includes a re-analysis of both the \kepler light curves and stellar parameters. We present atmospheric transmission spectra for both Kepler 51b and 51d in Section ~\ref{sec:results} while Section ~\ref{sec:discussion} discusses possible interpretations of the transmission spectra, the evolution of the Kepler 51 planets, and a comparison of these planets to other sub-Neptunes. We conclude in Section ~\ref{sec:conclusion}.}

\section{Observations} \label{sec:observation}

We observed two transits each of Kepler 51b and 51d with the infrared channel of WFC3 on \HST (Cycle 23, GO\#14218, PI Berta-Thompson). We used the G141 grism for all observations, providing slitless spectroscopy across the $1.1-1.7\mu m$ wavelength range \citep{dressel.2010.wfcihv}. We gathered continuous time-series spectra with the $256\times256$ subarray with {\tt SAMP-SEQ=SPARS10}, {\tt NSAMP=15} readout setting, resulting in exposure times of 103 seconds and on-target photon-collecting duty-cycle of 85\%. Given the relative faintness of Kepler 51 ($J=13.56$) compared to the other exoplanet host stars, we did not employ the commonly-used spatial scan technique \citep{mccullough.2012.cussww} as it would have resulted in a lower duty cycle or prohibitively high background counts.

The \HST visits were 8 and 12 orbits each for Kepler 51b and  51d, to provide sufficient out-of-transit baseline before and after transit. Direct images were taken for wavelength calibration at the start of each visit. To schedule the visits while accounting for the known TTVs in the Kepler 51 system, we provided STScI with a linear ephemeris (period and epoch) that we had fit to the transit timing predictions made by dynamical models for the years surrounding when \HST would likely observe. These timing predictions proved accurate, and \HST successfully observed transits on 29 September 2015 and 26 June 2016 for Kepler 51b and on 31 December 2015 and 24 January 2017 for Kepler 51d.

We extracted 1D spectra from the reduced {\tt FLT} images downloaded from MAST. These initial images are the result of the up-the-ramp fitting performed by the {\tt calwf3} pipeline, during which cosmic rays were removed and units were converted to photons/second. We used the custom IDL toolkit developed in \citet{berta.et.al.2012} for extraction. \added{We used the median of all science images to manually construct a mask that removed rectangular boxes around obvious stellar sources, and we calculated a background flux estimate for each exposure as the median of the unmasked pixels. We extracted the first-order spectrum from background-subtracted images using a rectangular aperture with a total height of 20 pixels, summing the flux vertically along each column because the tilt of the trace is small enough to be neglected. In addition to the background flux, we record the geometric properties of the spectral trace (centroid, slope, width) as diagnostic metrics for each exposure. No additional outliers were identified in the images or extracted spectra, indicating the {\tt calwf3} pipeline was effective at removing cosmic rays from these staring-mode data.}

In the first Kepler 51b visit, the first-order spectrum of a nearby star just barely overlaps with the blue wing of the Kepler 51 spectral trace. The first-order contamination is negligible at the wavelengths actually used in transit fits, but the alignment implies that the unseen second-order spectrum overlaps with Kepler 51's first order, potentially affecting the reddest third of the spectrum. Based on the first order spectra, we estimate this contaminating star to be roughly 10\% as bright as Kepler 51 at G141 wavelengths. The per-pixel flux from the G141 second order is less than $10\%$ of that from the first order, due both to its weak throughput ($5\times$ lower at its peak) and higher dispersion (roughly $2\times$ more disperse). Therefore, the star contributes about 1\% at most of the extracted flux. This contamination will dilute Kepler 51b's 0.5\% transit depth by 1\%, or about 50~ppm. We ignore this dilution, as it is significantly smaller than any of our achieved transit depth uncertainties.

\section{Analysis}\label{sec:analysis}

\subsection{HST White Light Curve Analysis} \label{wlc}

After extracting time-series stellar spectra, we created a white light curve of each visit by summing the wavelengths between 1.15 and 1.63~$\mu m$. We discarded the first\HST orbit from each visit due to the larger systematic variations compared to the other orbits \citep[e.g.][]{brown.et.al.2001,berta.et.al.2012,kreidberg.et.al.2014}
This left us with seven orbits per visit for Kepler 51b, and 11 orbits per visit for Kepler 51d. We calculated the uncertainties by combining the photon noise for each summed spectrum with the determined background noise for each point in time, and assume a Poisson distribution for both sources. We find an average uncertainty of 355~ppm for each point in the white light curve.


We construct a model flux time series $M = T \times S \times E$ which is the product of a {\tt batman} transit model ($T$) \citep{batman.paper}, an analytic systematics model ($S$) and an external parameter model ($E$). The equation for our systematics model is: $S = (C_{1}+C_{2}t_{v}+C_{3}t_{v}^{2})(1-Re^{-t_{b}/\tau})$ \citep{berta.et.al.2012, stevenson.et.al.2014}. Here $t_{v}$ and $t_{b}$ represent the time since the start of the visit and since the start of the \HST orbit respectively. Our free parameters are the coefficients $C_{1}$, $C_{2}$, $C_{3}$, the ramp amplitude $R$, and the ramp timescale $\tau$.

Our external parameter model ($E$) is a linear combination of the background value, and the spectral trace's slope, y centroid position, and width, all being functions of time ($E = 1+A_{\rm back}X_{\rm back}(t) + A_{\rm slope}X_{\rm slope}(t) + A_{\rm centroid}X_{\rm centroid}(t) + A_{\rm width}X_{\rm width}(t)  )$. As the x direction is wavelength, we do not include the x centroid as an external parameter. We standardized each of the four external parameters by subtracting the mean value and then dividing it by its standard deviation. Standardizing these parameters allowed us to compare the amplitude of their variations directly. We fitted for the amplitudes of these parameters ($A_{\rm back}, A_{\rm slope}, A_{\rm centroid}, A_{\rm width}$).

\deleted{The fitting routine consisted of two steps. We first fit the data with the above model via a least squares minimization for the $R_{p}/R_{s}$, mid-transit time ($T_{0}$), $a/R_{s}$, and the inclination ($i$). We assumed a zero eccentricity for each planet. These parameters combined with the systematics ($S$) and external parameter ($E$) models yielded a total of 13 fitted parameters. We used the $\chi^{2}_{r}$ from this initial fit of each transit to inflate the error bars (only if the $\chi^{2}_{r}$ is greater than 1) and repeated the procedure using a Markov Chain Monte Carlo method as implemented in the {\tt emcee} affine-invariant sampler \citep{mcmc.paper}. A uniform prior was assumed for all variables. We confirmed that the Markov chains converged for each visit by monitoring the Gelman-Rubin statistic for each parameter and by visually inspecting the trajectories of the walkers.}

\added{We fit the data with the above model using a Markov Chain Monte Carlo method as implemented in the {\tt emcee} affine-invariant sampler \citep{mcmc.paper} for $R_{p}/R_{s}$, mid-transit time ($T_{0}$), $a/R_{s}$, and the inclination ($i$). In order to properly represent the uncertainty on each point, we scale the error by a fitted parameter. Therefore, the final uncertainty for each point in our light curve is the original error multiplied by the point scaling term.  We initially assumed a zero eccentricity for each planet. We will discuss the validity of this assumption below. A uniform prior was assumed for all variables (14 in total). Using 100 walkers with 5000 steps (the initial 1000 were excluded as burn-in), we confirmed that the Markov chains converged for each visit by monitoring the Gelman-Rubin statistic so that every chain reaches $R <1.2$ \citep{brooks.and.gelman}.}

We employed a quadratic limb darkening law determined by the {\tt LDTK} package detailed in \citet{ldtk.paper} for a star with a temperature of 5673~K, $\log g$ of 4.697 and metallicity of 0.047 dex \citep{johnson.et.al.2017}. The coefficients, u=0.235 and v=0.195, are held constant throughout the white light curve fits. Relaxing this constraint left the parameters $a/R_{s}$ and inclination unconstrained even though the MCMC converged. We decided not to allow the limb darkening coefficients to vary as the gaps between \HST orbits leaves the shape of the transit to only be weakly constrained by the data, especially near ingress/egress. Our primary goal in fitting the white light curves is simply to determine the best-fit transit parameters for use in the spectroscopic light curve analysis (Section~\ref{sec:spectra_analysis}). We tested and confirmed that the mid-transit times do not depend sensitively on whether we vary the limb-darkening or keep it fixed.

\begin{figure}[h]
\includegraphics[width=10cm]{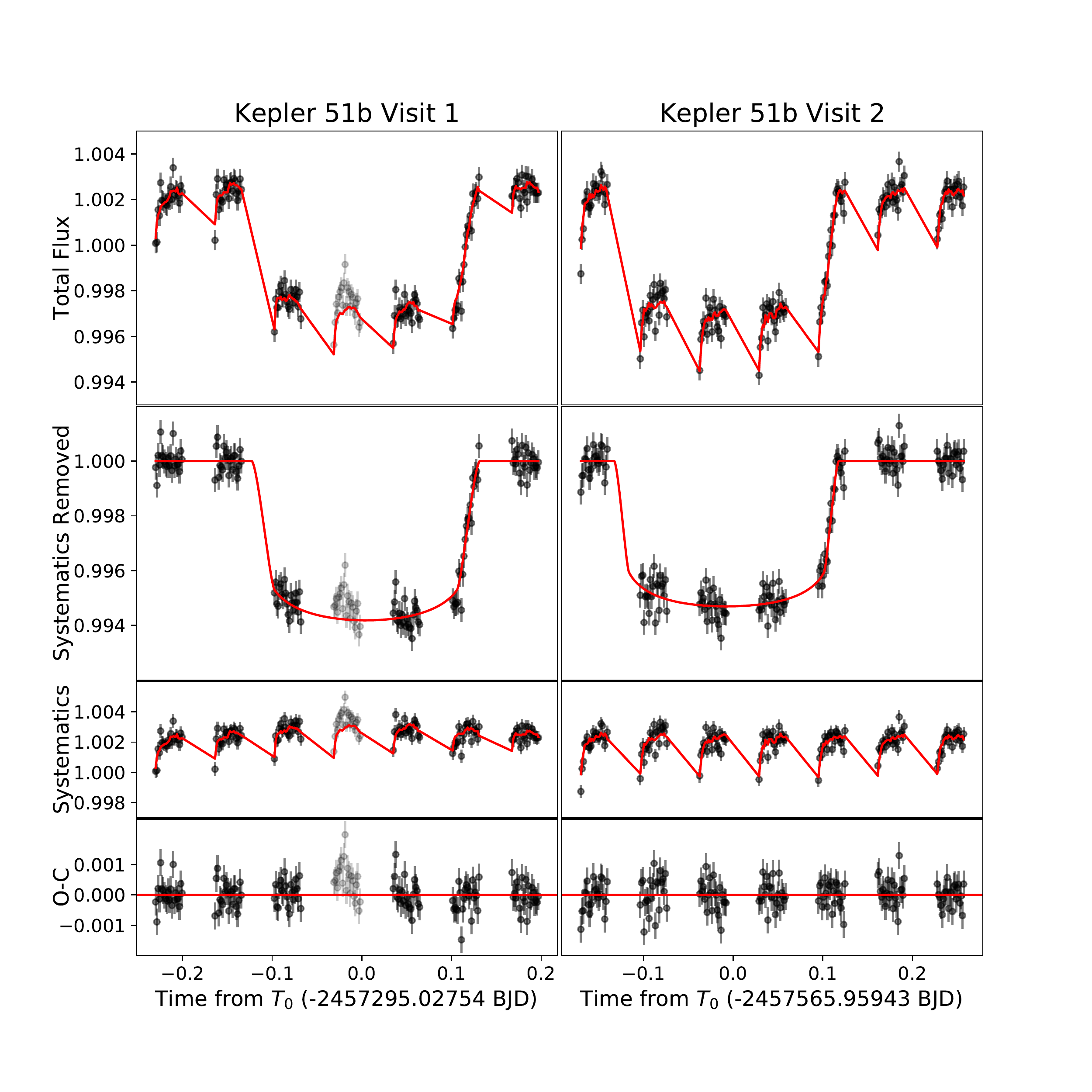}
\centering
\caption{The MCMC best-fit model of the white light curves of Kepler 51b in which we fit for both the transit ($T$) and the systematics model ($S \times E$) simultaneously. The top panel plots the combined systematics model and transit model to the data. The second panel is the transit minus the systematics while the third panel down is the combination of both the ramp-like systematics model and the external parameters. The bottom panel are the residuals from the fit. The data from the fourth \HST orbit of the first transit observation is apparently subject to a star-spot crossing event. We therefore removed this \HST orbit from future analysis. After binning the residuals, we determine that the model removes most of the correlated noise present.}
\label{fig:wlc_kep51b}
\end{figure}

\begin{figure}[h]
\includegraphics[width=10cm]{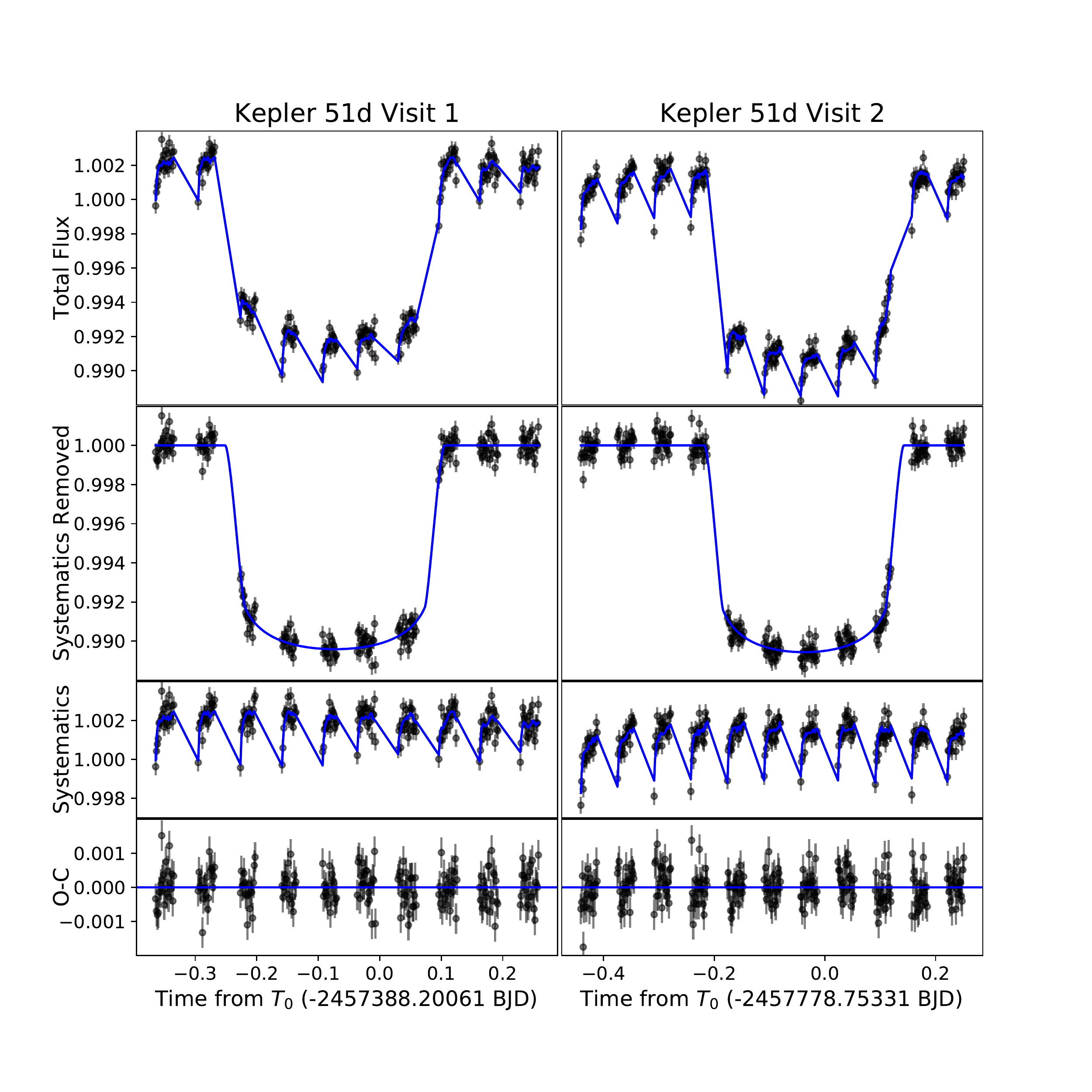}
\centering
\caption{The MCMC best-fit model of the white light curves of Kepler 51d in which we fit for both the transit and the systematics model simultaneously. Panels are the same as in Figure~\ref{fig:wlc_kep51b}. For Visit 1, we do not include the orbit demonstrating unusual systematic effects and only 10 orbits are shown. We again confirm that the systematics model captures and removes most of the red noise present.}
\label{fig:wlc_kep51d}
\end{figure}

The best-fit models are plotted in Figures~\ref{fig:wlc_kep51b} and~\ref{fig:wlc_kep51d} for Kepler 51b and 51d respectively. The top panel illustrates the full model. The second panel down isolates the transit model ($T$) by dividing out the systematics. The third panel isolates the instrumental effects ($S \times E$) by dividing out the transit, and the bottom panel are the residuals. The best-fit transit model parameters and their 1$\sigma$ uncertainties are listed in Table~\ref{tab:wlc_params}. For completeness, we also list the systematic parameters for each fit in Table~\ref{tab:wlc_sys}. 

\added{While most external parameters (background, y centroid, slope, width), are consistent with zero, we find that it is necessary to include these terms during our Kepler 51d Visit 1 fits. For the other visit of Kepler 51d and Kepler 51b, including this model makes very little difference on our resulting parameters and  our $\chi^{2}_{r}$ and BIC. However, we opt to include it for all our visits in order to remain consistent with our fits. We also note that \citet{agol.et.al.2010} found that including a quadratic term in time can bias the transit depth. However, we do not find this to be the case for any visits, and include it in our systematics model as it significantly minimizes both the $\chi^{2}_{r}$ and BIC for Kepler 51d Visit 2. The inclusion of this term had less than a 1$\sigma$ effect on all parameter values for the other visits.}

\added{We also test the strength of our results by allowing the eccentricity and argument of periastron to vary. We place a Gaussian prior on these two terms using our results from our TTV analysis (Section~\ref{sec:ttv}). Not only do we recover the same values as our previous fits, but find that we also return the priors for both eccentricity and argument of periastron. We therefore conclude that our data is insensitive to both terms, and holding these constant has no effect on our end model.}

During the first Kepler 51b visit, we noted a star-spot crossing event during the fourth \HST orbit and therefore removed this orbit from the fit. We also found that the second \HST orbit (the first orbit after the removal of the large systematic orbit) in the first visit of Kepler 51d demonstrated unusual systematics, which we were unable to correct with the ramp-like model. We removed this orbit, which improved the $\chi_{r}^{2}$ by 20\%.

\begin{table*}[]
\centering
\caption{The best-fit parameters for each white light curve for both Kepler 51b and Kepler 51d. We report asymmetric uncertainties when the upper and lower 1$\sigma$ error bars do not agree to within 10\% of each other.}
\label{tab:wlc_params}
\resizebox{\textwidth}{!}{
\begin{tabular}{l|c|c|c|c|}
\cline{2-5}
                      & \multicolumn{2}{c|}{Kepler 51b}                                        & \multicolumn{2}{c|}{Kepler 51d}                                        \\ \cline{2-5} 
                      & Visit 1                               & Visit 2                        & Visit 1                               & Visit 2                        \\ \hline
${R_{p}}/{R_{s}}$ & $0.0725 \pm 0.0014$        & $0.0689 \pm 0.0013$ & $0.0969\pm0.0011$        & $0.0976\pm 0.0009$ \\
$T_{0}$ ($BJD_{TDB}$) & $2457295.03189^{+0.00171}_{-0.00286}$ & $2457565.95248 \pm 0.00398$    & $2457388.20063^{+0.00099}_{-0.00143}$ & $2457778.75336 \pm 0.00136$    \\
${a}/{R_{s}}$     & $57.4 \pm 3.7$               & $59.7 \pm 2.9$        & $123.5^{+3.4}_{-8.1}$              & $124.9^{+2.2}_{-5.4}$       \\
$i$ ($^{\circ}$) (e=0)         & $89.62^{+0.07}_{-0.06}$            & $89.78^{+0.15}_{-0.17}$     & $89.88^{+0.08}_{-0.10}$            & $89.91^{+0.06}_{-0.08}$     \\
$b$ & $0.38^{+0.12}_{-0.20}$ & $0.22 \pm 0.16$ & $0.25 \pm 0.18$ & $0.19^{+0.16}_{-0.13}$  \\
Point Scaling & 0.94 & 1.03 & 0.98 & 1.06  \\
$\chi^{2}_{r}$ (dof)  & 1.01 (132)                           & 1.13 (154)                    & 1.35 (204)                           & 1.22 (240)                   
\end{tabular}
}
\end{table*}

\begin{table*}[]
\centering
\caption{The best-fit parameters for the systematics model ($S\times E$).}
\label{tab:wlc_sys}
\resizebox{0.9\textwidth}{!}{
\begin{tabular}{lcccc}
\cline{2-5}
\multicolumn{1}{l|}{}        & \multicolumn{2}{c|}{Kepler 51b}                                                                          & \multicolumn{2}{c|}{Kepler 51d}                                                           \\ \cline{2-5} 
\multicolumn{1}{l|}{}        & \multicolumn{1}{c|}{Visit 1}                              & \multicolumn{1}{c|}{Visit 2}                 & \multicolumn{1}{c|}{Visit 1}                & \multicolumn{1}{c|}{Visit 2}                                     \\ \hline
\multicolumn{1}{l|}{$C_{1}$}     & \multicolumn{1}{c|}{$1.0025 \pm 0.0004$}                 & \multicolumn{1}{c|}{$1.0030 \pm 0.0005$}     & \multicolumn{1}{c|}{$1.0030 \pm 0.0002$}    & \multicolumn{1}{c|}{$0.9984 \pm 0.0009$}                         \\
\multicolumn{1}{l|}{$C_{2}$} & \multicolumn{1}{c|}{$(4.15 \pm 2.56)\times 10^{-3}$}               & \multicolumn{1}{c|}{$(-2.15 \pm 2.44)\times 10^{-3}$} & \multicolumn{1}{c|}{$(-2.98 \pm 2.01)\times 10^{-3}$} & \multicolumn{1}{c|}{$(1.14 \pm 2.95)\times 10^{-3}$}                      \\
\multicolumn{1}{l|}{$C_{3}$} & \multicolumn{1}{c|}{$(-1.05 \pm 0.52)\times10^{-3}$}              & \multicolumn{1}{c|}{$(-2.89 \pm 4.79)\times 10^{-3}$} & \multicolumn{1}{c|}{$(1.12 \pm 2.69)\times 10^{-3}$} & \multicolumn{1}{c|}{$(-5.53 \pm 1.74)\times 10^{-3}$}                     \\
\multicolumn{1}{l|}{$R$}     & \multicolumn{1}{c|}{$(1.51 \pm 0.95)\times 10^{-3}$}               & \multicolumn{1}{c|}{$(3.18 \pm 0.76)\times 10^{-3}$}  & \multicolumn{1}{c|}{$(2.46 \pm 0.25) \times 10^{-3})$} & \multicolumn{1}{c|}{$(2.70 \pm 0.28)\times 10^{-3}$}                      \\

\multicolumn{1}{l|}{$\tau$}  & \multicolumn{1}{c|}{$(3.27 \pm 4.46)\times 10^{-3}$}               & \multicolumn{1}{c|}{$(3.33 \pm 0.77)\times 10^{-3}$}  & \multicolumn{1}{c|}{$(1.89 \pm 0.38)\times 10^{-3}$} & \multicolumn{1}{c|}{$(2.77 \pm 0.62)\times 10^{-3}$}                      \\
Background                 & \multicolumn{1}{|c|}{$(-4.17 \pm 2.82)\times 10^{-4}$}               & \multicolumn{1}{c|}{$(-5.17 \pm 4.36)\times 10^{-5}$}  & \multicolumn{1}{c|}{$(3.02 \pm 0.70)\times 10^{-4}$}  & \multicolumn{1}{c|}{$(1.29 \pm 0.61)\times 10^{-4}$}  \\
Y Centroid                 & \multicolumn{1}{|c|}{$0.26 \pm 2.45)\times 10^{-4}$}                & \multicolumn{1}{c|}{$(6.13 \pm 2.24)\times 10^{-4}$}   & \multicolumn{1}{c|}{$(-1.54 \pm 0.69)\times 10^{-4}$} & \multicolumn{1}{c|}{$(-7.15 \pm 4.16)\times 10^{-4}$} \\
Slope                      & \multicolumn{1}{|c|}{$(6.69 \pm 8.81)\times 10^{-5}$}                & \multicolumn{1}{|c|}{$(1.07 \pm 0.42)\times 10^{-3}$}   & \multicolumn{1}{c|}{$(-2.04 \pm 0.69)\times 10^{-4}$} & \multicolumn{1}{c|}{$(-1.95 \pm 0.79)\times 10^{-3}$} \\
Width                      & \multicolumn{1}{|c|}{$(5.19 \pm 1.58)\times 10^{-4}$} & \multicolumn{1}{c|}{$(1.087 \pm 2.25)\times 10^{-4}$}   & \multicolumn{1}{c|}{$(-1.29 \pm 0.81)\times 10^{-4}$} & \multicolumn{1}{c|}{$(-2.80 \pm 1.40)\times 10^{-4}$}
\end{tabular}
}
\end{table*}

The difference in radius ratios between the two visits for Kepler 51b varied by less than 2$\sigma$, and the radius ratios for Kepler 51d by less than $1\sigma$. It is possible the slight variations in transit depth between the two Kepler 51b transits are partially due to stellar activity. This slight variation in transit depth was also observed in the \kepler light curves, and are discussed in more detail in Section~\ref{sec:kepler_compare}. The other transit parameters demonstrate less than a 1$\sigma$ deviation between the two visits of each planet.

\subsection{HST Spectroscopic Light Curve Analysis}\label{sec:spectra_analysis}

\begin{table}

\movetabledown=2in

\begin{rotatetable}

\caption{WFC3 Spectroscopic Light Curve Data}

{\footnotesize

\begin{tabular}{c|c|c|c|c|c|c|c|c|c|c|c}
BJD$_{\rm TDB}$\tablenotemark{a} & 
$f_{1.152-1.176}$\tablenotemark{b} & 
$\sigma_{f, 1.152-1.176}$\tablenotemark{c}  & 
...\tablenotemark{d}  & 
$f_{1.602-1.625}$ \tablenotemark{b}& 
$\sigma_{f, 1.602-1.625}$ \tablenotemark{c} &  
$X_{\rm background}$\tablenotemark{e} & 
$X_{\rm centroid}$\tablenotemark{f} &
$X_{\rm slope}$\tablenotemark{g} & 
$X_{\rm width}$\tablenotemark{h} & 
Planet & 
Visit \\

\hline \hline

2457294.797597 & 0.997822 & 0.002080 & ...  & 0.998915 & 0.002225  & 1.608083 & -0.133704 & 0.009397 & 0.752109 & b & 1 \\
2457294.798998 & 0.998306 & 0.002076 & ...  & 1.002302 & 0.002223  & 1.465450 & -0.134871 & 0.009368 & 0.749086 & b & 1 \\
2457294.800398 & 1.002192 & 0.002074 & ...  & 0.997460 & 0.002211  & 1.325458 & -0.130375 & 0.009359 & 0.746755 & b & 1 \\
... & ... & ... & ...  & ...  & ...  & ... & ... & ... & ... & ... & ... \\
2457779.037859 & 0.997767 & 0.002077 & ...  & 1.002081 & 0.002242  & 1.334146 & 0.250196 & 0.008083 & 0.741572 & d & 2 \\
2457779.039260 & 0.998326 & 0.002077 & ...  & 1.003961 & 0.002242  & 1.333878 & 0.246535 & 0.008146 & 0.742031 & d & 2 \\
2457779.040660 & 1.003909 & 0.002082 & ...  & 1.003823 & 0.002242  & 1.347367 & 0.237440 & 0.008232 & 0.741259 & d & 2 \\

\end{tabular}

\tablecomments{Table \ref{tab:wfc3_data} is published in its entirety as a machine-readable table. A portion is shown here for guidance regarding its form and content.}

\tablenotetext{a}{The mid-exposure time.}
\tablenotetext{b}{The relative flux within a wavelength bin, with subscripts representing the wavelength bounds (in $\mu$m). }
\tablenotetext{c}{The uncertainty on the relative flux within the corresponding wavelength bin.}
\tablenotetext{d}{In the full table, each wavelength is represented by two columns: flux $f_\lambda$ and flux uncertainty $\sigma_{f, \lambda}$.} 
\tablenotetext{e}{The sky background, in photoelectrons/pixel/s.} 
\tablenotetext{f}{The vertical centroid of the spectral trace on the detector (pixels).} 
\tablenotetext{g}{The slope of the spectral trace on the detector (pixels/pixel).} 
\tablenotetext{h}{The width of the spectral trace on the detector (pixels).}

}

\label{tab:wfc3_data}

\end{rotatetable}
\end{table}

We applied the divide-white approach similar to the method discussed in \citet{kreidberg.et.al.2014} to create the spectroscopic light curves. To start, we summed the spectroscopic data into 5 pixel bins, which corresponds to an approximate 25~nm wavelength span. This allowed us to generate 20 spectroscopic light curves, all of which are presented in machine-readable format in Table \ref{tab:wfc3_data}. We isolated the systematic noise time series by first dividing out the white light curve transit model (panel 2 in Figures~\ref{fig:wlc_kep51b} and~\ref{fig:wlc_kep51d}) from the data. This left us with the residuals: a combination of noise, the ramp-like model, and effects from the external parameters. These residuals were then divided out from every spectroscopic light curve. With this method, we made the assumption that the systematic noise for the light curves was  wavelength-independent across all wavelength bins. We supported this assumption by noting that the $\chi_{r}^{2}$ for each spectroscopic fit was near 1 and that the residuals binned down with the expected Gaussian noise trend of $1/\sqrt{N}$.

\deleted{We performed the same fitting routine as the white light curve fits using both a least squares minimization and a MCMC.} \added{We performed the same fitting routine as the white light curve fits by running an MCMC and scaling the uncertainties by a fitted parameter.} We fixed the $a/R_{s}$ and inclination to the values determined from the \kepler light curves reported in \citet{masuda.2014}. \added{We find less than a 2$\sigma$ deviation between the \citet{masuda.2014} and our \HST values for these parameters. However, as the \kepler light curves have a well-sampled ingress and egress we used the more precise values derived from the \kepler data set for this analysis}. We set the mid-transit time constant to the best-fit values determined from the \HST white light curve. The derived $R_{p}/R_{s}$ values for each of the 20 spectroscopic light curves are listed in Tables~\ref{tab:spectra_kep51b} and~\ref{tab:spectra_kep51d} for Kepler 51b and 51d respectively. These tables also include the assumed quadratic limb darkening coefficients, again determined using {\tt LDTK} and held constant in the fits. We found that a four-parameter non-linear limb darkening law used by \citet{ldtk.paper} had no effect on the transit depth. However, applying a non-linear limb darkening law significantly decreased the efficiency for light curve fitting. Thus we opted to apply a quadratic limb darkening law for each spectroscopic light curve. \replaced{We also modeled a vertical offset and visit-long slope for each spectroscopic light curve. While the systematic residuals captured most instrumental noise, there remained a slight slope in the data that appeared to depend on wavelength.}{We also fit for and removed a visit-long slope for each spectroscopic light curve. The slope and offset varied by wavelength, but including these parameters in our fit improved both the BIC and $\chi^{2}_{r}$ for all spectroscopic light curves during each visit.} 

\begin{table*}[]
\centering
\caption{Wavelength ranges, limb darkening parameters, and planetary radii for Kepler 51b.}
\begin{tabular}{cc|cc|cc}
\multicolumn{1}{l}{}                                                  & \multicolumn{1}{l|}{}                                              & \multicolumn{2}{c|}{Visit 1}          & \multicolumn{2}{c|}{Visit 2}          \\ \hline
 $\lambda (\mu m)$ & Limb Darkening $(u,v)$ & $R_{p}/R_{s}$        & $\chi^{2}_{r}$ & $R_{p}/R_{s}$        & $\chi^{2}_{r}$ \\ \hline
\multicolumn{1}{c|}{1.152 - 1.176}                                    & (0.247, 0.267)                                                     & $0.0751 \pm 0.0026$ & 1.03          & $0.0681 \pm 0.0027$ & 1.06          \\
\multicolumn{1}{c|}{1.176 - 1.200}                                    & (0.240, 0.271)                                                     & $0.0728 \pm 0.0025$ & 0.95         & $0.0661 \pm 0.0027$ & 0.94         \\
\multicolumn{1}{c|}{1.200 - 1.223}                                    & (0.235, 0.273)                                                     & $0.0721 \pm 0.0026$ & 0.91         & $0.0729 \pm 0.0024$ & 1.09          \\
\multicolumn{1}{c|}{1.223 - 1.247}                                    & (0.227, 0.287)                                                     & $0.0780 \pm 0.0025$ & 1.16          & $0.0647 \pm 0.0027$ & 1.06          \\
\multicolumn{1}{c|}{1.247 - 1.271}                                    & (0.223, 0.288)                                                     & $0.0720 \pm 0.0025$ & 0.96         & $0.0728 \pm 0.0023$ & 1.06          \\
\multicolumn{1}{c|}{1.271 - 1.294}                                    & (0.187, 0.319)                                                     & $0.0723 \pm 0.0027$ & 1.18          & $0.0694 \pm 0.0024$ & 1.03          \\
\multicolumn{1}{c|}{1.294 - 1.318}                                    & (0.209, 0.308)                                                     & $0.0719 \pm 0.0025$ & 0.96         & $0.0677 \pm 0.0025$ & 1.07          \\
\multicolumn{1}{c|}{1.318 - 1.342}                                    & (0.248, 0.192)                                                     & $0.0709 \pm 0.0025$ & 0.96         & $0.0680 \pm 0.0022$ & 0.93         \\
\multicolumn{1}{c|}{1.342 - 1.366}                                    & (0.240, 0.198)                                                     & $0.0688 \pm 0.0024$ & 0.88         & $0.0717 \pm 0.0020$ & 0.78         \\
\multicolumn{1}{c|}{1.366 - 1.389}                                    & (0.233, 0.201)                                                     & $0.0727 \pm 0.0020$ & 0.98         & $0.0717 \pm 0.0024$ & 1.11          \\
\multicolumn{1}{c|}{1.389 - 1.413}                                    & (0.226, 0.207)                                                     & $0.0725 \pm 0.0027$ & 1.07          & $0.0720 \pm 0.0023$ & 1.04          \\
\multicolumn{1}{c|}{1.413 - 1.437}                                    & (0.222, 0.209)                                                     & $0.0732 \pm 0.0026$ & 1.05          & $0.0672 \pm 0.0024$ & 0.98         \\
\multicolumn{1}{c|}{1.437 - 1.460}                                    & (0.211, 0.211)                                                     & $0.0701 \pm 0.0026$ & 0.91         & $0.0697 \pm 0.0026$ & 1.19          \\
\multicolumn{1}{c|}{1.460 - 1.484}                                    & (0.202, 0.218)                                                     & $0.0690 \pm 0.0027$ & 1.04          & $0.0660 \pm 0.0025$ & 0.96         \\
\multicolumn{1}{c|}{1.484 - 1.508}                                    & (0.199, 0.213)                                                     & $0.0693 \pm 0.0024$ & 0.7553         & $0.0668 \pm 0.0026$ & 1.004          \\
\multicolumn{1}{c|}{1.508 - 1.531}                                    & (0.188, 0.223)                                                     & $0.0746 \pm 0.0026$ & 0.98         & $0.0671 \pm 0.0031$ & 1.45          \\
\multicolumn{1}{c|}{1.531 - 1.555}                                   & (0.179, 0.221)                                                     & $0.0663 \pm 0.0032$ & 1.12          & $0.0734 \pm 0.0025$ & 1.04          \\
\multicolumn{1}{c|}{1.555 - 1.579}                                    & (0.174, 0.220)                                                     & $0.0749 \pm 0.0027$ & 0.95         & $0.0707 \pm 0.0024$ & 0.95         \\
\multicolumn{1}{c|}{1.579 - 1.602}                                    & (0.174, 0.211)                                                     & $0.0698 \pm 0.0029$ & 1.05          & $0.0644 \pm 0.0027$ & 0.96         \\
\multicolumn{1}{c|}{1.602 - 1.625}                                    & (0.165, 0.211)                                                     & $0.0682 \pm 0.0033$ & 1.04          & $0.0676 \pm 0.0028$ & 1.05         
\end{tabular}
\label{tab:spectra_kep51b}
\end{table*}

\begin{table*}[]
\centering
\caption{Wavelength ranges, limb darkening parameters, and planetary radii for Kepler 51d.}
\begin{tabular}{cc|cc|cc}
\multicolumn{1}{l}{}                                                  & \multicolumn{1}{l|}{}                                              & \multicolumn{2}{c|}{Visit 1}          & \multicolumn{2}{c|}{Visit 2}          \\ \hline
 $\lambda (\mu m)$ & Limb Darkening $(u,v)$ & $R_{p}/R_{s}$        & $\chi^{2}_{r}$ & $R_{p}/R_{s}$        & $\chi^{2}_{r}$ \\ \hline
\multicolumn{1}{c|}{1.152 - 1.176}                                   & (0.247, 0.267)                                                     & $0.096 \pm 0.0014$ & 0.98         & $0.0976\pm0.0014$ & 1.01          \\
\multicolumn{1}{c|}{1.176 - 1.200}                                    & (0.240, 0.271)                                                     & $0.0971\pm0.0014$   & 1.00         & $0.0985\pm0.0014$ & 1.09          \\
\multicolumn{1}{c|}{1.200 - 1.223}                                    & (0.235, 0.273)                                                     & $0.0985\pm0.0014$   & 1.01          & $0.0996\pm0.0013$ & 1.05          \\
\multicolumn{1}{c|}{1.223 - 1.247}                                    & (0.227, 0.287)                                                     & $0.0954\pm0.0014$   & 0.95         & $0.0972\pm0.0013$ & 0.96         \\
\multicolumn{1}{c|}{1.247 - 1.271}                                    & (0.223, 0.288)                                                     & $0.0948\pm0.0013$   & 0.96         & $.1000\pm0.0013$  & 1.01          \\
\multicolumn{1}{c|}{1.271 - 1.294}                                    & (0.187, 0.319)                                                     & $0.0961\pm0.0014$   & 0.98         & $0.0961\pm0.0013$ & 0.95         \\
\multicolumn{1}{c|}{1.294 - 1.318}                                    & (0.209, 0.308)                                                     & $0.0938\pm0.0014$   & 1.01          & $0.0985\pm0.0012$ & 1.01          \\
\multicolumn{1}{c|}{1.318 - 1.342}                                    & (0.248, 0.192)                                                     & $0.0975\pm0.0013$   & 0.95         & $0.0987\pm0.0012$ & 1.01          \\
\multicolumn{1}{c|}{1.342 - 1.366}                                    & (0.240, 0.198)                                                     & $0.0960\pm0.0014$   & 1.11          & $0.0989\pm0.0013$ & 0.84         \\
\multicolumn{1}{c|}{1.366 - 1.389}                                    & (0.233, 0.201)                                                     & $0.0951\pm0.0015$   & 1.18          & $0.0971\pm0.0012$ & 1.01          \\
\multicolumn{1}{c|}{1.389 - 1.413}                                    & (0.226, 0.207)                                                     & $0.0962\pm0.0014$   & 1.16          & $0.0992\pm0.0011$ & 0.86         \\
\multicolumn{1}{c|}{1.413 - 1.437}                                    & (0.222, 0.209)                                                     & $0.0965\pm0.0015$   & 1.12          & $0.0988\pm0.0013$ & 0.92         \\
\multicolumn{1}{c|}{1.437 - 1.460}                                    & (0.211, 0.211)                                                     & $0.0964\pm0.0014$   & 1.14          & $0.0957\pm0.0014$ & 1.09          \\
\multicolumn{1}{c|}{1.460 - 1.484}                                    & (0.202, 0.218)                                                     & $0.0960\pm0.0015$   & 1.00         & $0.0948\pm0.0014$ & 1.01          \\
\multicolumn{1}{c|}{1.484 - 1.508}                                    & (0.199, 0.213)                                                     & $0.0941\pm0.0014$   & 0.92         & $0.0976\pm0.0013$ & 1.03          \\
\multicolumn{1}{c|}{1.508 - 1.531}                                    & (0.188, 0.223)                                                     & $0.0979\pm0.0013$   & 0.78         & $0.0977\pm0.0014$ & 1.03          \\
\multicolumn{1}{c|}{1.531 - 1.555}                                   & (0.179, 0.221)                                                     & $0.0961\pm0.0015$   & 0.99         & $0.0956\pm0.0014$ & 1.00         \\
\multicolumn{1}{c|}{1.555 - 1.579}                                    & (0.174, 0.220)                                                     & $0.0977\pm0.0014$   & 0.97         & $0.0966\pm0.0014$ & 1.12          \\
\multicolumn{1}{c|}{1.579 - 1.602}                                    & (0.174, 0.211)                                                     & $0.0960\pm0.0016$   & 1.05          & $0.0965\pm0.0014$ & 0.98         \\
\multicolumn{1}{c|}{1.602 - 1.625}                                    & (0.165, 0.211)                                                     & $0.0973\pm0.0015$   & 0.95         & $0.0970\pm0.0014$ & 0.89        
\end{tabular}
\label{tab:spectra_kep51d}
\end{table*}

\begin{figure*}[h]
\includegraphics[width=\textwidth]{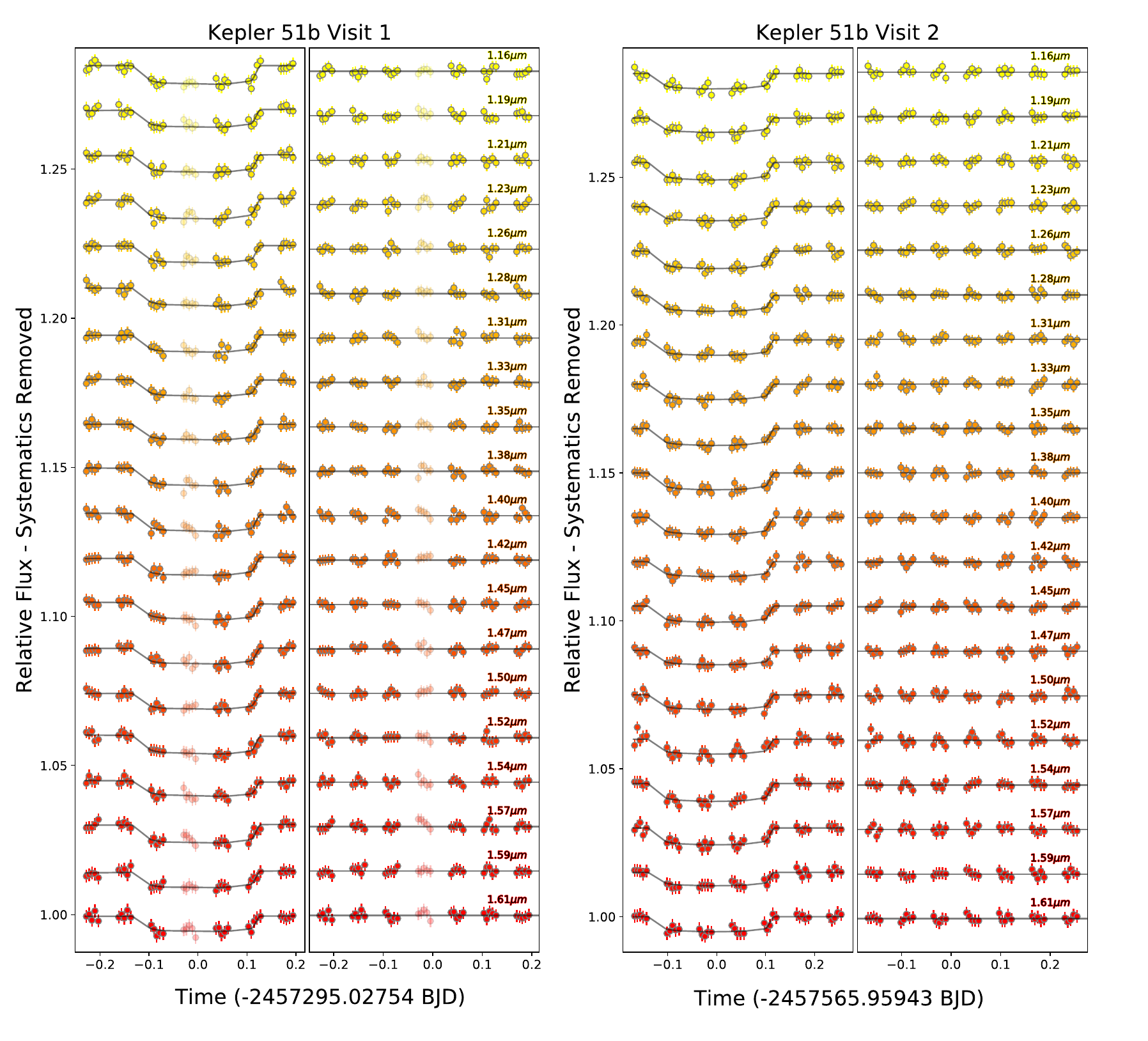}
\caption{Spectroscopic light curves from the two Kepler 51b visits. For clarity, we bin each orbit to just 5 points (4 exposures per bin) and offset each transit along the y axis. Each light curve represents a wavelength bin of approximately 25~nm with the shortest wavelength bin at the top and moving downwards to sequentially larger wavelengths, as labeled to the right. The time is represented by an offset from the expected mid-transit time assuming a linear ephemeris. We do not include the \HST orbit with the star-spot crossing event in the model for the Kepler 51b Visit 1. The residuals for each fit are plotted to the right of their respective transit. All sets of residuals demonstrate Gaussian noise when binned.}
\label{fig:spectra_curves_kep51b}
\end{figure*}

\begin{figure*}[h]
\includegraphics[width=\textwidth]{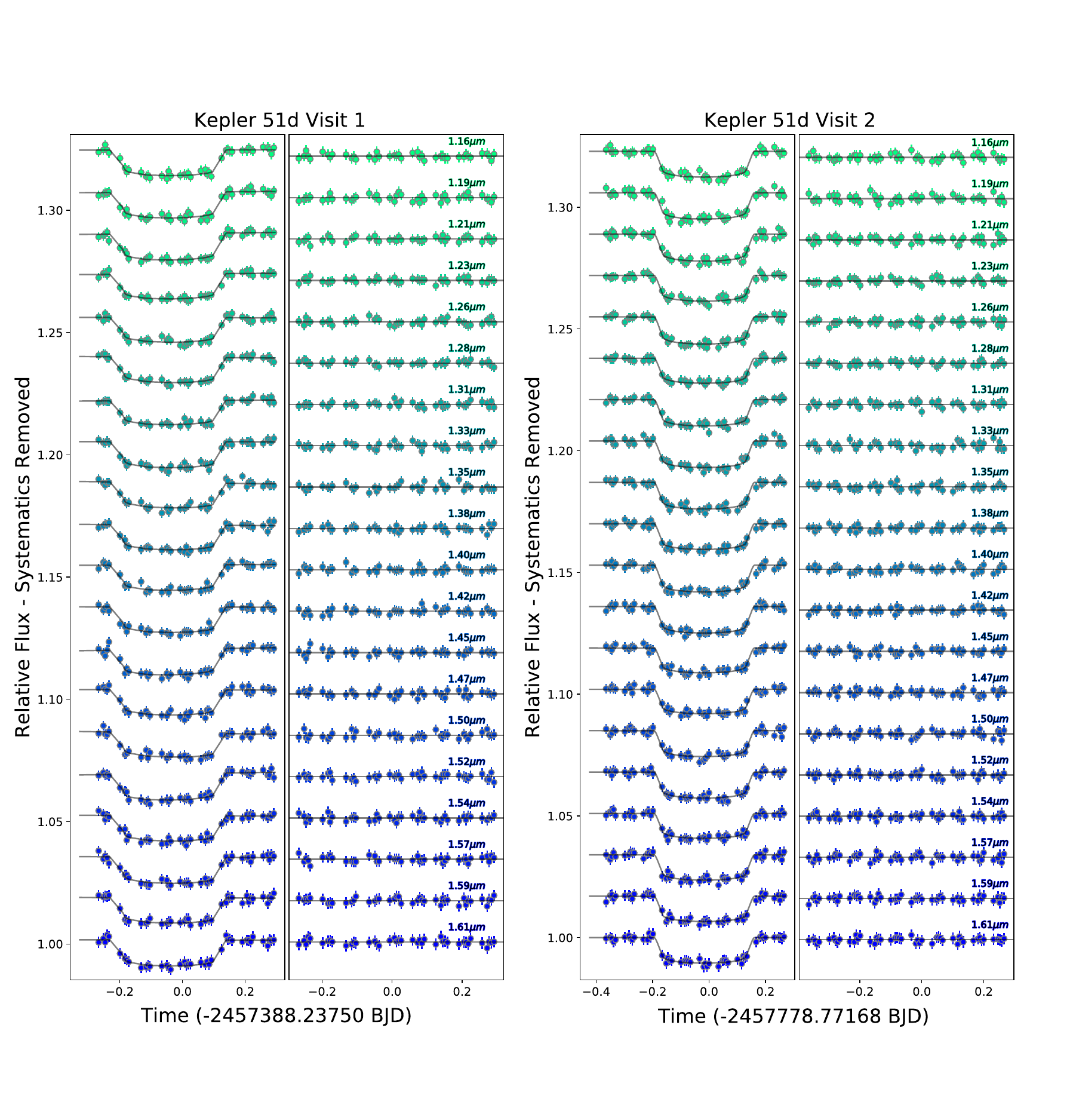}
\caption{Spectroscopic light curves from the two Kepler 51d visits, similar to those shown in Figure~\ref{fig:spectra_curves_kep51b}. As with the spectroscopic light curves of Kepler 51b, all residuals demonstrate Gaussian noise when binned down.}
\label{fig:spectra_curves_kep51d}
\end{figure*}

Figures~\ref{fig:spectra_curves_kep51b} and ~\ref{fig:spectra_curves_kep51d} highlight the spectroscopic light curves (minus the systematics) for Kepler 51b and 51d respectively. We include the corresponding residuals to the right of each transit. \added{By binning the residuals, we find that the noise scales as expected for independent Gaussian-distributed measurements.} The mid-wavelength value is noted to the right of the residuals and each wavelength bin spans approximately 25~nm. Each spectroscopic light curve is binned and offset slightly in these figures for clarity.

\subsection{Re-Analysis of the Kepler Data}\label{sec:kepler_analysis}

TTVs of the Kepler 51 system were first reported in \citet{steffen.et.al.2013} and later analyzed by \citet{masuda.2014} who determined all three planets had extremely low densities. To provide additional confirmation of the unusually low planet masses, we independently re-derived transit times from the full \kepler dataset. To do this, we isolated each transit of Kepler 51b, 51c, and 51d for all \kepler quarters. We included a two-day baseline (one day on either side) for each transit light curve. We fit for a third order polynomial \added{in time} as well as the $a/R_{s}$, inclination, transit depth, and mid-transit time for each transit for both the long cadence (Q1-11) and short cadence data (Q12-16). \citet{masuda.2014} discovered a double transit of Kepler 51b and 51d in the data, which also included a likely star-spot crossing event. Because of the complicated nature of this event, we do not include this transit in our analysis.

We first assumed the limb darkening coefficients reported in \citet{masuda.2014} and then repeated the analysis by using the limb darkening coefficients determined by {\tt LDTK}. We found a less than 1$\sigma$ variation in all parameters between the two limb darkening representations. We therefore decided to \replaced{adopted}{adopt} the \citet{masuda.2014} \kepler quadratic limb darkening parameters for the rest of the analysis.

Using the MCMC method, we modeled each individual transit of Kepler 51b, 51c, and 51d independently. We find that while long cadence data is adequate for constraining the mid-transit times, it struggled to precisely constrain the transit depth, $a/R_{s}$, and inclination for individual planetary transits. We discuss these parameters further in Section~\ref{sec:kepler_compare}. \added{We report our mid-transit times for Kepler 51b, 51c, and 51d in Tables ~\ref{tab:51btimes}, ~\ref{tab:51ctimes}, and ~\ref{tab:51dtimes} respectively.}

\begin{table*}[]
\centering
\caption{Mid-transit times for Kepler 51b including those determined from the \HST observed transits.}
\begin{tabular}{ccccc}
Epoch & \begin{tabular}[c]{@{}c@{}}$t_{0}$\\ (BJD -2454833)\end{tabular} & $1\sigma_{lower}$ & $1\sigma_{upper}$ & $\chi^{2}_{r}$ \\ \hline
0     & 159.11020                                                        & 0.00117           & 0.00138           & 1.38           \\
1     & 204.26341                                                        & 0.00116           & 0.00094           & 1.24           \\
2     & 249.41429                                                        & 0.00148           & 0.00165           & 1.32           \\
3     & 294.57387                                                        & 0.00182           & 0.00180           & 0.98           \\
4     & 339.72659                                                        & 0.00157           & 0.00168           & 0.98           \\
5     & 384.87849                                                        & 0.00109           & 0.00126           & 1.28           \\
6     & 430.03431                                                        & 0.00105           & 0.00103           & 0.89           \\
8     & 520.34296                                                        & 0.00140           & 0.00150           & 0.86           \\
9     & 565.50058                                                        & 0.00171           & 0.00169           & 1.67           \\
10    & 610.65750                                                        & 0.00116           & 0.00126           & 0.80           \\
11    & 655.81281                                                        & 0.00107           & 0.00130           & 1.17           \\
12    & 700.97563                                                        & 0.00187           & 0.00208           & 2.78           \\
13    & 746.12681                                                        & 0.00126           & 0.00153           & 0.84           \\
14    & 791.28725                                                        & 0.00158           & 0.00170           & 0.75           \\
15    & 836.44031                                                        & 0.00200           & 0.00214           & 1.12           \\
16    & 881.59877                                                        & 0.00096           & 0.00096           & 0.82           \\
17    & 926.75401                                                        & 0.00231           & 0.00165           & 1.35           \\
18    & 971.90560                                                        & 0.00154           & 0.00141           & 1.28           \\
19    & 1017.06093                                                       & 0.00192           & 0.00165           & 1.01           \\
20    & 1062.21173                                                       & 0.00171           & 0.00188           & 1.98           \\
21    & 1107.36679                                                       & 0.00101           & 0.00105           & 0.92           \\
22    & 1152.52080                                                       & 0.00086           & 0.00087           & 1.00           \\
23    & 1197.67694                                                       & 0.00089           & 0.00088           & 0.93           \\
24    & 1242.83033                                                       & 0.00091           & 0.00095           & 0.99           \\
25    & 1287.98438                                                       & 0.00086           & 0.00087           & 0.93           \\
26    & 1333.14164                                                       & 0.00089           & 0.00089           & 1.03           \\
27    & 1378.29828                                                       & 0.00090           & 0.00088           & 0.96           \\
28    & 1423.45432                                                       & 0.00091           & 0.00088           & 1.03           \\
29    & 1468.61332                                                       & 0.00088           & 0.00088           & 1.02           \\
51    & 2462.03211                                                       & 0.00203           & 0.00149           & 1.01           \\
57    & 2732.95267                                                       & 0.00303           & 0.00286           & 1.13          
\end{tabular}
\label{tab:51btimes}
\end{table*}

\begin{table*}[]
\caption{Mid-transit times for Kepler 51c}
\centering
\begin{tabular}{ccccc}
Epoch & \begin{tabular}[c]{@{}c@{}}$t_{0}$\\ (BJD - 2454833)\end{tabular} & $1\sigma_{lower}$ & $1\sigma_{upper}$ & $\chi^{2}_{r}$ \\ \hline
0     & 295.31635                                                         & 0.00393           & 0.00420           & 0.86           \\
1     & 380.63913                                                         & 0.00289           & 0.00300           & 1.09           \\
2     & 465.95364                                                         & 0.00271           & 0.00260           & 1.05           \\
3     & 551.26314                                                         & 0.00410           & 0.00406           & 1.26           \\
4     & 636.56959                                                         & 0.00353           & 0.00345           & 1.31           \\
7     & 892.51915                                                         & 0.00380           & 0.00382           & 1.23           \\
8     & 977.84155                                                         & 0.00444           & 0.00489           & 1.38           \\
10    & 1148.46496                                                        & 0.00438           & 0.00442           & 1.02           \\
11    & 1233.80861                                                        & 0.00315           & 0.00296           & 0.94           \\
12    & 1319.11300                                                        & 0.00420           & 0.00521           & 1.03           \\
14    & 1489.75234                                                        & 0.00333           & 0.00368           & 0.95          
\end{tabular}
\label{tab:51ctimes}
\end{table*}

\begin{table*}[]
\centering
\caption{Mid-transit times for Kepler 51d including those determined from the \HST observed transits.}
\begin{tabular}{ccccc}
Epoch & \begin{tabular}[c]{@{}c@{}}$t_{0}$\\ ($BJD -2454833$)\end{tabular} & $1\sigma_{lower}$ & $1\sigma_{upper}$ & $\chi^{2}_{r}$ \\ \hline
0     & 212.02419                                                        & 0.00086           & 0.00087           & 1.34           \\
1     & 342.20768                                                        & 0.00087           & 0.00086           & 1.27           \\
2     & 472.39081                                                        & 0.00097           & 0.00099           & 1.84           \\
3     & 602.57368                                                        & 0.00085           & 0.00085           & 1.39           \\
5     & 862.93204                                                        & 0.00121           & 0.00120           & 2.23           \\
6     & 993.10486                                                        & 0.00094           & 0.00092           & 1.45           \\
7     & 1123.28462                                                       & 0.00075           & 0.00073           & 1.12           \\
8     & 1253.45065                                                       & 0.00135           & 0.00133           & 2.07           \\
9     & 1383.63021                                                       & 0.00065           & 0.00066           & 0.93           \\
18    & 2555.20049                                                       & 0.00150           & 0.00099           & 1.35           \\
21    & 2945.75330                                                       & 0.00107           & 0.00139           & 1.22          
\end{tabular}
\label{tab:51dtimes}
\end{table*}

\subsection{Transit Timing Analysis}\label{sec:ttv}

From our re-analysis of the \kepler light curves, we confirm the mid-transit times listed in \citet{masuda.2014} for all three planets. \added{As we include a scaling relation on our uncertainties, we find that our mid-transit times are less well-constrained as those reported in \citet{masuda.2014}.} Combining these \added{newly fitted} times with the new \HST mid-transit times for Kepler 51b and 51d, we modeled the TTVs for all three planets using the {\tt TTVFast} code \citep{ttvfast}. For each planet, we fitted planet-to-star mass ratio, orbital period $P$, eccentricity and argument of periastron ($\sqrt{e}\cos\omega$ and $\sqrt{e}\sin\omega$), and time $t_0$ of inferior conjunction closest to the dynamical epoch $t_{\rm epoch}(\mathrm{BJD}-2454833)=150$. 
\replaced{These are osculating Jacobian orbital elements defined at $t_{\rm epoch}$, and the time $t_0$ is related to the time of periastron passage $\tau$ via $2\pi (t_0-\tau)/P = E_0 - e\sin E_0$, where $E_0=2\arctan\left[\sqrt{{1-e}\over{1+e}}\tan\left({\pi \over 4}-{\omega \over 2}\right)\right]$.}{Here $P$, $e$, and $\omega$ are osculating Jacobian orbital elements defined at $t_{\rm epoch}$. The time $t_0$ is converted to the time of periastron passage $\tau$ via $2\pi (t_0-\tau)/P = E_0 - e\sin E_0$, where $E_0=2\arctan\left[\sqrt{{1-e}\over{1+e}}\tan\left({\pi \over 4}-{\omega \over 2}\right)\right]$, and this $\tau$ is used to set the mean anomaly at time $t_{\rm epoch}$.}
The inclination and longitude of ascending node were fixed to be \replaced{$\pi/2$ and $0$}{$90^\circ$ and $0^\circ$}, respectively\replaced{.}{; this is justified because the mutual orbital inclinations of the three transiting planets are most likely small, and because inclination effects appear only in the higher order terms of the TTV signal \citep[e.g.,][]{2014ApJ...790...58N}.}
We adopted uniform priors for all these parameters and used {\tt emcee} \citep{mcmc.paper} to sample from their posterior distribution. We include the results from our TTV modeling in a machine-readable table. A sample of the posterior results as well as information on each parameter is shown in Table~\ref{tab:ttv_posteriors}.
\replaced{}{As discussed in detail by \citet{2016ApJ...818..177A} and \cite{2017AJ....154....5H}, the TTVs of planets c and d exhibit clear chopping signals, which break the mass--eccentricity degeneracy in the sinusoidal component of the TTV signal \citep{2012ApJ...761..122L} and fix the masses of these planets. Therefore the solution in this system is well constrained and appropriate for sampling with an MCMC.}


\begin{table}

\movetabledown=2in

\begin{rotatetable}

\caption{MCMC samples from our TTV analysis.}

{\tiny

\begin{tabular}{c|c|c|c|c|c|c|c|c|c|c|c|c|c|c}
\hline

$M_{p}/M_{s}$\tablenotemark{a} & 
$P$\tablenotemark{b}  & 
$\sqrt{e} \cos{ \omega }$\tablenotemark{c} &
$\sqrt{e} \sin{ \omega }$\tablenotemark{c} & 
$t_{0}$\tablenotemark{d}  & 
$M_{p}/M_{s}$\tablenotemark{a} & 
$P$\tablenotemark{b} &
$\sqrt{e} \cos{ \omega}$\tablenotemark{c} & 
$\sqrt{e}\sin{(\omega)}$\tablenotemark{c} & 
$t_{0}$\tablenotemark{d} & 
$M_{p}/M_{s}$\tablenotemark{a} &
$P$\tablenotemark{b} & 
$\sqrt{e}\cos{(\omega)}$\tablenotemark{c} & 
$\sqrt{e}\sin{(\omega)}$\tablenotemark{c}  & 
$t_{0}$\tablenotemark{d}  \\ \hline

(K51b) & 
(K51b) & 
(K51b) & 
(K51b) & 
(K51b) & 
(K51c) & 
(K51c) & 
(K51c) & 
(K51c) & 
(K51c) & 
(K51d) & 
(K51d) & 
(K51d) & 
(K51d) & 
(K51d) \\

\hline \hline
$1.31\times10^{-5}$ & 45.1542      & -0.12                          & -0.16                          & 159.111       & $1.17\times10^{-5}$ & 85.3143      & -0.09                          & -0.06                          & 295.321       & $1.58\times10^{-5}$ & 130.1852     & -0.10                          & -0.05                          & 212.024       \\
$1.48\times10^{-5}$ & 45.1542      & -0.12                          & -0.10                          & 159.110       & $1.39\times10^{-5}$ & 85.3144      & 0.04                           & 0.02                           & 295.321       & $1.84\times10^{-5}$ & 130.1856     & -0.04                          & 0.03                           & 212.023       \\
$1.78\times10^{-5}$ & 45.1540      & -0.13                          & -0.15                          & 159.111       & $1.30\times10^{-5}$ & 85.3170      & 0.04                           & -0.03                          & 295.318       & $1.64\times10^{-5}$ & 130.1853     & -0.06                          & 0.00                           & 212.023       \\
$1.01\times10^{-5}$ & 45.1544      & -0.09                          & -0.08                          & 159.111       & $1.54\times10^{-5}$ & 85.3128      & 0.02                           & -0.05                          & 295.322       & $1.70\times10^{-5}$ & 130.1851     & -0.05                          & -0.02                          & 212.023      \\
... & ...      &...                        &...                       & ...    & ... & ...    & ...                         & ...                         & ...       &... & ...   & ...                        & ...                         & ...     
\end{tabular}

\tablecomments{Table \ref{tab:ttv_posteriors} is published in its entirety as a machine-readable table. A portion is shown here for guidance regarding its form and content.}

\tablenotetext{a}{The ratio of the mass of the planet to the mass of the star (unitless).}
\tablenotetext{b}{The period of the planet (days).}
\tablenotetext{c}{In these columns, $e$ is the eccentricity of the planet and $\omega$ the argument of periastron.}
\tablenotetext{d}{Time of first inferior conjunction assuming a Keplerian orbit (BJD-2454833).}

}

\label{tab:ttv_posteriors}

\end{rotatetable}
\end{table}

\begin{figure}[h]
\includegraphics[width=0.33\textwidth]{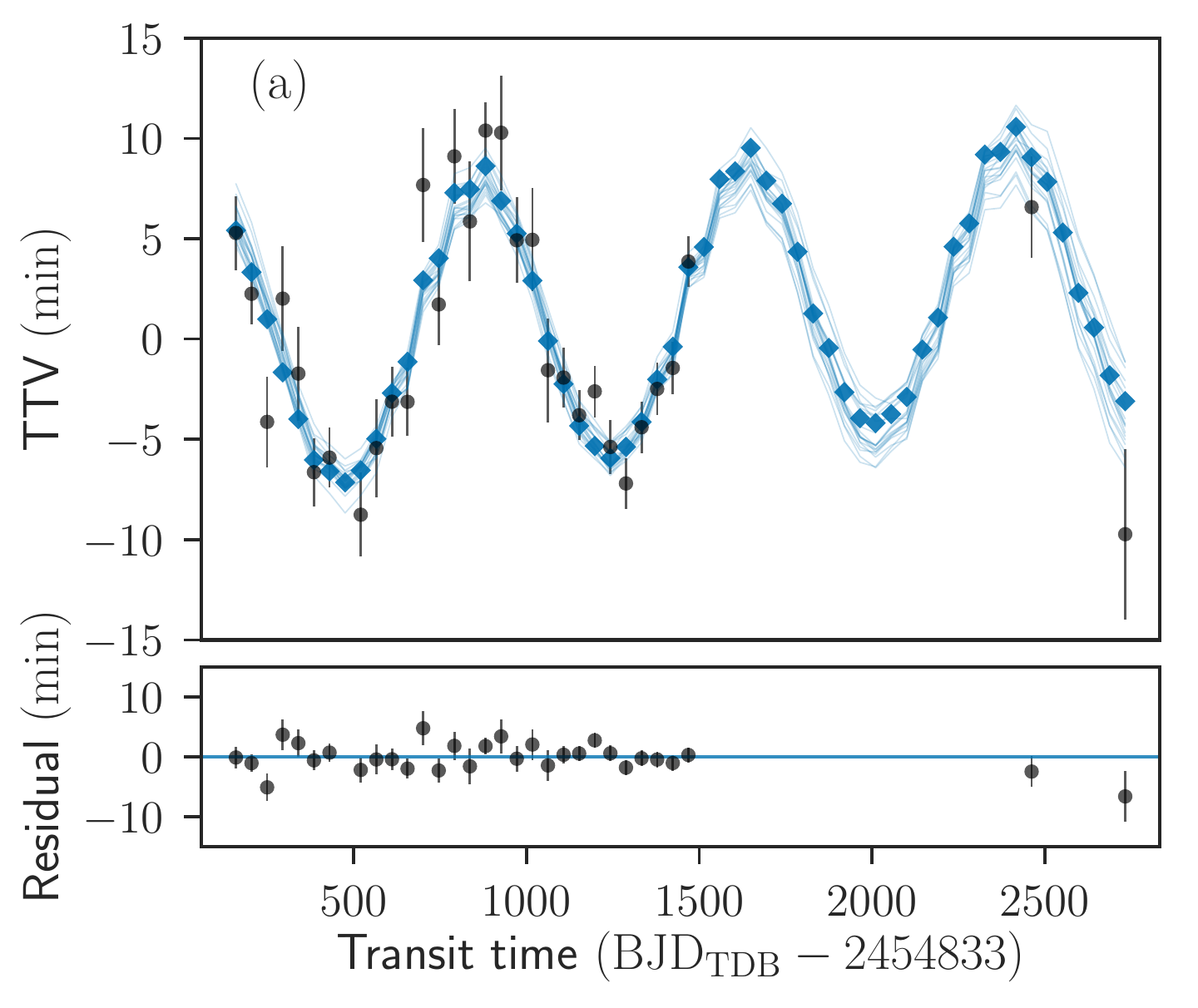}
\includegraphics[width=0.33\textwidth]{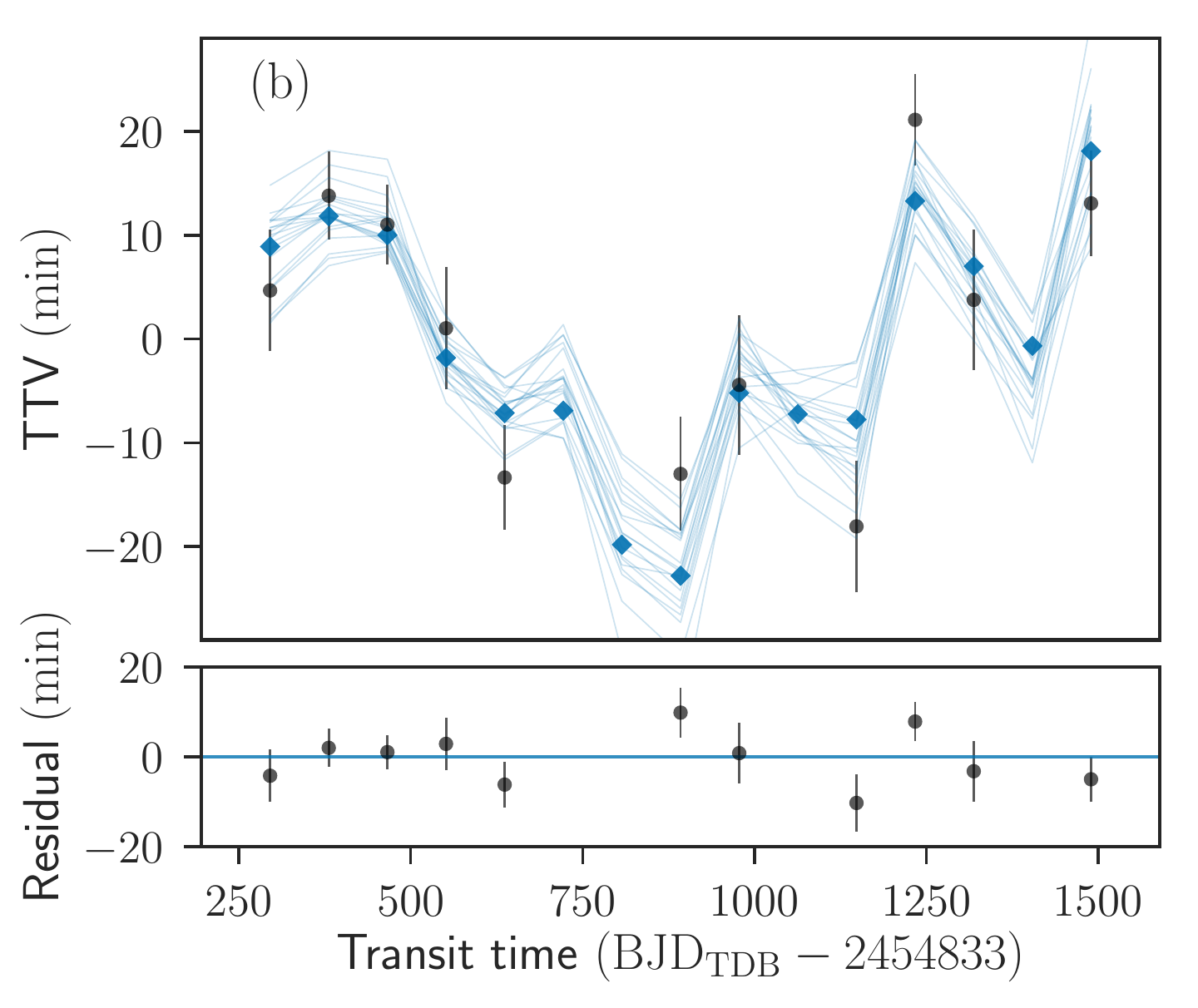}
\includegraphics[width=0.33\textwidth]{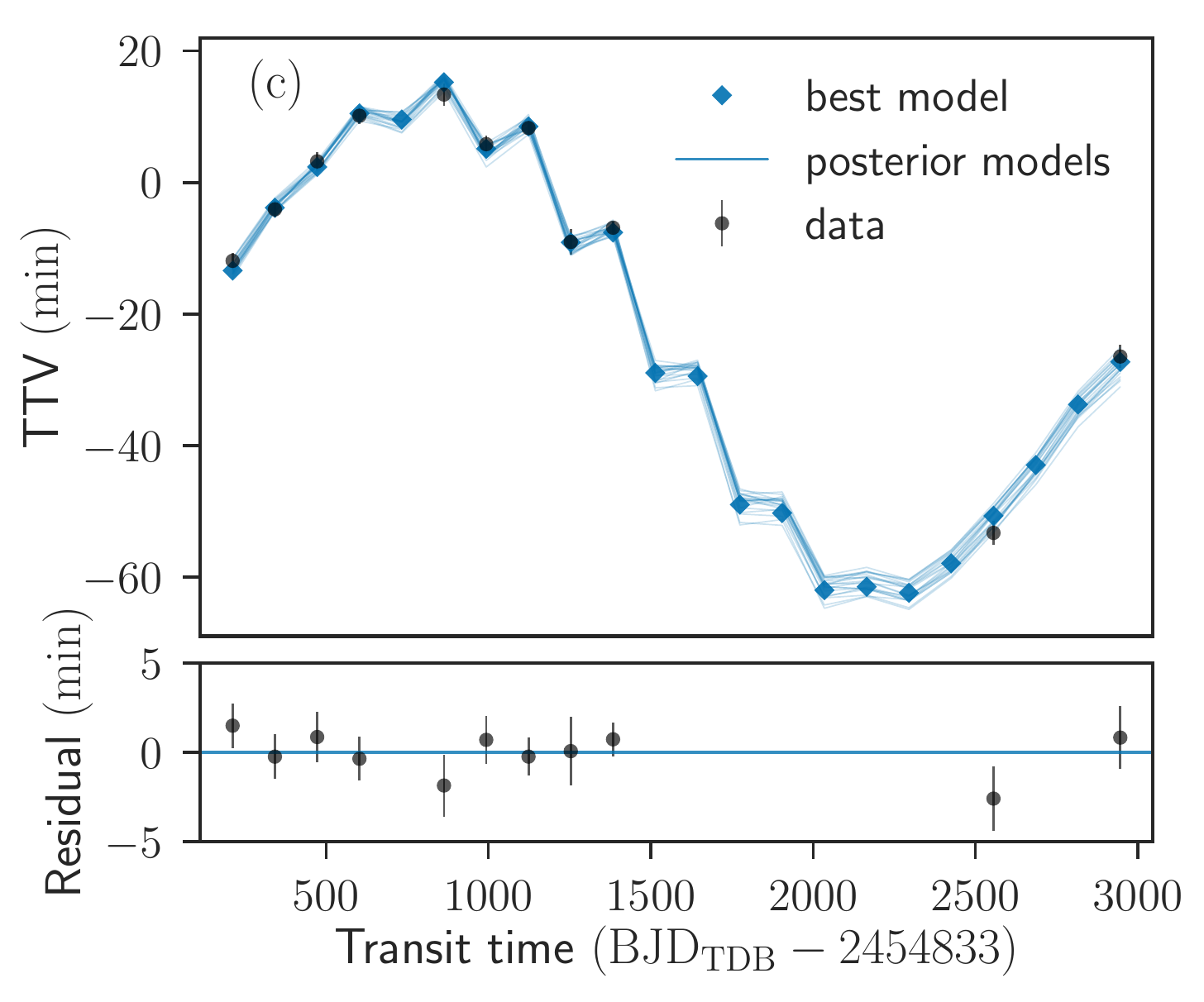}
\caption{Transit timing variations for Kepler 51b (a), 51c (b), and 51d (c), determined from both \kepler and \HST observations. The best-fit model is plotted with blue points and we include random posterior models in light blue in order to demonstrate the spread of possibilities. \deleted{By including the \HST mid-transit times, we are better able to constrain the mass ratios for all three planets compared to those presented in \citet{masuda.2014}.} }
\label{fig:ttv}
\end{figure}

We plot the determined TTVs in Figure~\ref{fig:ttv} along with models generated with parameters drawn from the inferred posterior distribution. Because Kepler 51c is a grazing transit, it has the largest uncertainties on the mid-transit time, which leads to the largest spread in models. The updated planetary parameters determined by this TTV analysis are listed in Table~\ref{tab:ttv_params}. \replaced{In particular, by adding the \HST mid-transit times to those from \kepler, we updated the planet-to-star mass ratios $M_{p}/M_{s}$ between  \citet{masuda.2014} and this work from $(0.61^{+0.43}_{-0.23})\times10^{-5}$ \replaced{in }{}to $(1.13 \pm 0.56)\times10^{-5}$ for Kepler 51b, from $(1.15 \pm0.12)\times10^{-5}$ to $(1.35 \pm 0.16) \times 10^{-5}$ for Kepler 51c, and  from $(2.19\pm0.32)\times10^{-5}$ to $(1.74 \pm 0.35) \times 10^{-5}$ for Kepler 51d.} {We find that our \HST times are both consistent to the previous model based solely on \kepler mid-transit times while also allowing a better constraint on the planet-to-star mass ratios of the three planets. However, we still find our mass-ratio values less constrained when compared to \citet{masuda.2014}. We attribute this discrepancy to the larger mid-transit times we report, a consequence from allowing the size of our uncertainties on the flux to change during initial fitting.}

\begin{table*}[]
\caption{Planetary parameters for Kepler 51b, 51c, and 51d, determined from the TTV analysis of both \kepler and \HST mid-transit times}. $t_{0}$ represents the time of the first inferior conjunction assuming Keplerian orbits, not the mid-transit time. 

\begin{tabular}{lccc}

\multicolumn{1}{c}{}                                                           & Kepler 51b                                     & Kepler 51c                                   & Kepler 51d              \\ \hline
\multicolumn{1}{l|}{$\frac{M_{p}}{M_{s}}$} & \multicolumn{1}{c|}{$(1.13 \pm 0.56) \times 10^{-5}$} & \multicolumn{1}{c|}{$(1.35 \pm 0.16) \times 10^{-5}$}         & $(1.74 \pm 0.35) \times 10^{-5}$         \\
\multicolumn{1}{l|}{$P$ (days)}                                             & \multicolumn{1}{c|}{$45.1542 \pm 0.0003$}  & \multicolumn{1}{c|}{$85.3139\pm0.0017$}  & $130.1845\pm0.0007$ \\
\multicolumn{1}{l|}{$e$}                                    & \multicolumn{1}{c|}{$0.03 \pm 0.01$}      & \multicolumn{1}{c|}{$0.01^{+0.02}_{-0.01}$}       & $0.01 \pm 0.01$       \\
\multicolumn{1}{l|}{$\omega (^{\circ})$}                                    & \multicolumn{1}{c|}{$53.3 \pm 14.6$}       & \multicolumn{1}{c|}{$-16.7^{+65.9}_{-44.4}$}      & $-12.6^{+68.2}_{-50.6}$      \\
\multicolumn{1}{l|}{$t_{0} (-2454833 BJD)$}                                    & \multicolumn{1}{c|}{$159.1105\pm0.0005$}   & \multicolumn{1}{c|}{$295.3202\pm0.0026$} & $212.0234\pm0.0007$

\end{tabular}
\label{tab:ttv_params}
\end{table*}

\subsection{Kepler 51 Stellar Parameters Analysis}\label{sec:stellar_analysis}

Previously published work on Kepler 51 \citep{steffen.et.al.2013, masuda.2014} relied on the KIC stellar parameters for Kepler 51 \citep{brown.et.al.2011}, which contain large uncertainties. Here, we better constrain the estimates of the mass and radius of Kepler 51 by combining a high resolution spectrum of Kepler 51 from the California Kepler Survey \citep{johnson.et.al.2017}, parallax measurements from Gaia DR2 \citep{gaia.et.al.2016, gaia.et.al.2018}, and a gyrochronological age from \kepler \citep{mcquillan.et.al.2013, angus.et.al.2015}. 

\citet{mcquillan.et.al.2013} determined the rotation period of Kepler 51 to be $8.222\pm0.007$ days. Using gyrochronology relations in \citet{angus.et.al.2015}, we used this rotation period to determine an approximate age of 500 Myr for this system. \citet{angus.et.al.2015} also notes that different gyrochronology relations yield systematic age uncertainties at the level of 35-50\%. Therefore, we assumed a cautious error bar of 50\% or $\pm$ 250 Myr uncertainty on the age of Kepler 51. Kepler 51 rotates faster than stars of similar $B-V$ color in the Hyades cluster, which has a cited age of 625 Myr \citep{perryman.et.al.1998}. We therefore conclude that the estimated age of 500 Myr for Kepler 51 is reasonable. However, \citet{gaia.et.al.2018b} estimates an older age of 790 Myr for the Hyades cluster, but this 20\% increase in age is still smaller than the 50\% uncertainty we have assumed for Kepler 51. 

Kepler 51 is also an active star, which supports the argument for its young age. We note that approximately 17\% of the short cadence light curves for Kepler 51b and 51d show by-eye signs of a star-spot crossing event. We also observed a star-spot crossing in the Kepler 51b Visit 1 observation. \added{\citet{mcquillan.et.al.2013} determined the activity of Kepler 51 had an amplitude of 11.95 mmags in the Kepler bandpass. Using the correlation of amplitude to spot coverage in \citet{rackham.et.al.2019}, we find that this must correlate to an average spot coverage of 4-6\% for Kepler 51, depending on whether faculae is included in the model. We leave the detailed modeling of this stellar activity for future analysis.}

We used the {\tt isochrones} modeling package \citep{morton.2015} to determine the mass and radius posteriors of Kepler 51. We used the {\tt emcee} fitting option, and place Gaussian priors on the distance, effective temperature, metallicity, and $\log g$ as listed in Table~\ref{tab:stellar_params}, as well as published stellar magnitudes from the KIC \citep{brown.et.al.2011}. We determined a mass of $0.985 \pm 0.012 M_{\odot}$ and a radius of $0.881 \pm 0.011 R_{\odot}$ for Kepler 51. These values are relatively insensitive to the age prior: changing center of the prior from 500 Myr to 750 Myr has basically no effect. In fact, if we keep {\tt isochrones}' default uninformed power law age prior, we determine a best-fit mass of $0.972 \pm 0.021 M_{\odot}$ and radius of $0.886 \pm 0.015  R_{\odot}$. These values correspond to a less than a 1$\sigma$ difference between the informed and uninformed age priors. With the unconstrained power law age prior, {\tt isochrones}  determines an approximate age of $1.8_{-1.15}^{+1.3}$ Gyr for Kepler 51, which is in agreement (albeit less precise) with that determined by gyrochronology.

\added{\citet{fulton.et.al.2018} performed a similar fit for Kepler 51's mass and radius using the high resolution data combined with Gaia astrometry. We find less than a $1\sigma$ deviation between the two radii values; however, our mass is about $1.5\sigma$ larger than their reported value. We attribute this to both the inclusion of a younger age (stars become brighter as they age) as well as potentially our choice in stellar models. \citet{fulton.et.al.2018} use MIST stellar evolution models while we opted for the Dartmouth stellar evolution models, both built into the {\tt isochrones} framework. However, we later find that this slight discrepancy in the mass is small in comparison to the mass ratios of each planet and thus has minimal effect on our final planetary masses.}

\begin{figure}[h]
\includegraphics[width=8cm]{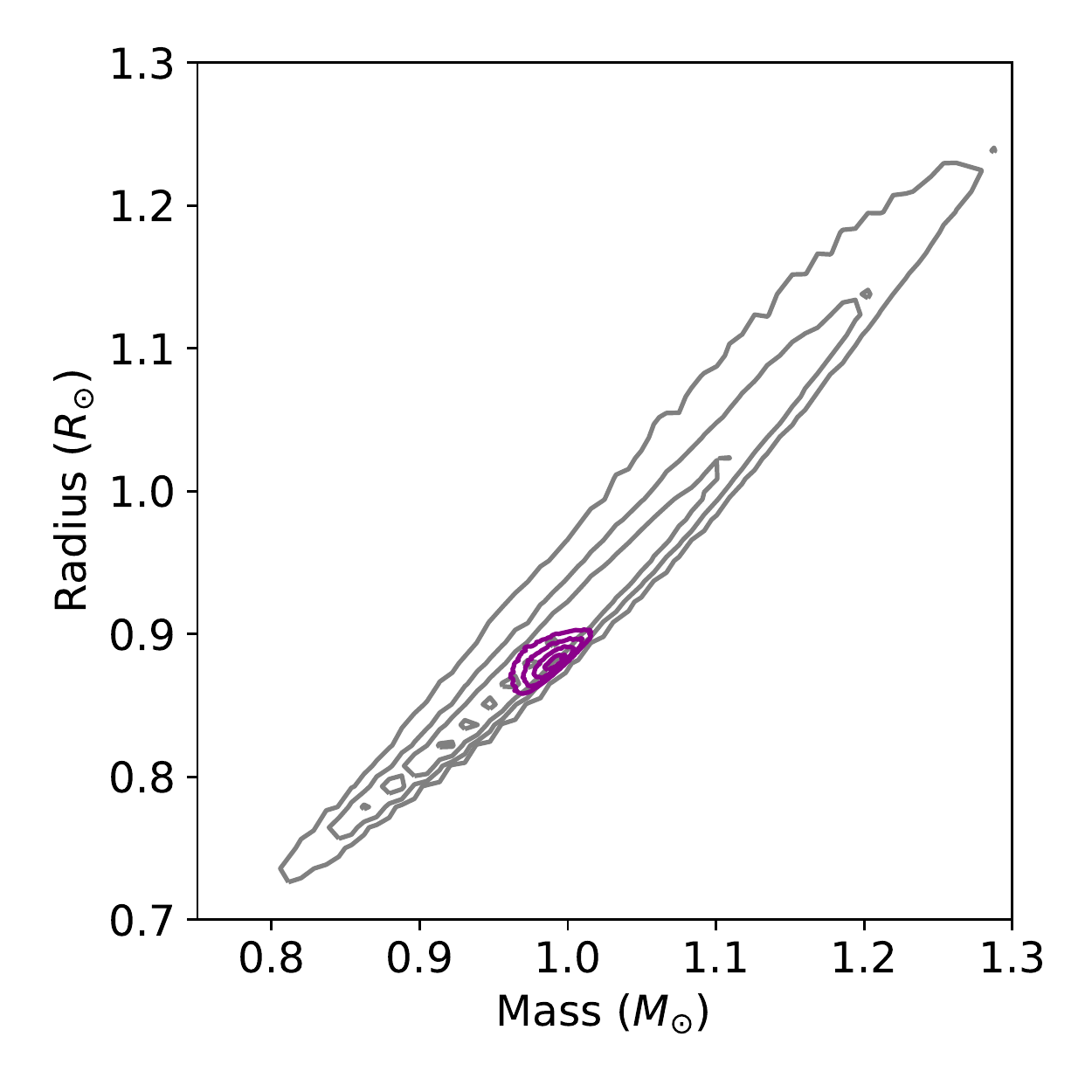}
\caption{The joint posteriors determined by {\tt isochrones} for the mass and radius of Kepler 51 using photometry alone (gray contours) and photometry plus high-resolution spectroscopy, the Gaia DR2 distance measurement, and age priors (purple contours). Contours for both models are the 1, 2, and 3$\sigma$ levels. We find that our updated stellar mass and radius are in good agreement with values determined by photometry alone, and much better constrained. Based on the mass and radius information, we conclude that Kepler 51 is a young solar-type star.}
\label{fig:mass_rad_post}
\end{figure}

We re-ran {\tt isochrones} with only the published photometry values for Kepler 51, which was all the information available to constrain the mass/radius of Kepler 51 in the KIC. The 1, 2, and 3$\sigma$ uncertainty contours of this photometry-only model are plotted in gray in Figure ~\ref{fig:mass_rad_post}. For comparison, we included the same contour levels from the model which incorporates all currently available information listed in Table~\ref{tab:stellar_params}. With the added information, we both agree with the mass/radius KIC values cited in \citet{masuda.2014} while much more precisely constraining the mass and radius for this star.

We acknowledge that a 2\% constraint in the stellar mass and radius is very precise, even with the new information provided by high resolution spectroscopy, gyrochronology and the Gaia DR2 parallax. We do not account for systematic uncertainties that might have been generated by the {\tt isochrones} fitting routine and hence the uncertainties on the stellar mass and radius may be slightly under-approximated. 

\begin{table*}[]
\centering
\caption{Stellar parameters for Kepler 51, including their corresponding references. We determine the mass, radius, and density for Kepler 51 using the {\tt isochrones} package from \citet{morton.2015}.}
\begin{tabular}{lcc}
\multicolumn{3}{c}{Kepler 51 Stellar Parameters}                                                                  \\ \hline
\multicolumn{1}{l|}{log(g)}                  & $4.7\pm0.1$                 & Johnson et al. 2017                \\
\multicolumn{1}{l|}{\replaced{$Z_{metal}$}{$Z$} (dex)}       & $0.05\pm0.04$                & Johnson et al. 2017                \\
\multicolumn{1}{l|}{$T_{eff}$ (K)}           & $5670\pm60$) & Johnson et al. 2017                \\
\multicolumn{1}{l|}{$P_{rot}$ (days)}        & $8.222\pm0.007$               & McQuillan et al. 2013              \\
\multicolumn{1}{l|}{Age (Myr)}               & $500\pm250$                   & This Work                          \\

\multicolumn{1}{l|}{Distance (pc)}           & $802\pm14$                    & Gaia et al. 2018 \\
\multicolumn{1}{l|}{$M_{s}$ ($M_{\odot}$)}   & $0.985\pm0.012$               & This Work                          \\
\multicolumn{1}{l|}{$R_{s}$ ($R_{\odot}$)}   & $0.881\pm0.011$               & This Work                          \\
\multicolumn{1}{l|}{$\rho_{s}$ ($\mathrm{g/cm^{3}}$)} & $2.03\pm0.08$                 & This Work                         
\end{tabular}
\label{tab:stellar_params}
\end{table*}

\subsection{Updated Planetary Masses and Radii}
\label{sec:mr}

We apply the updated stellar parameters to the mass ratios determined from the TTV analysis and radius ratios determined from the \HST and \kepler transits in order to verify the extreme low-densities of all three Kepler 51 super-puff planets. We do this by first randomly sampling from the stellar posterior in order to determine joint mass and radius values for Kepler 51. We repeat this random sampling of the mass ratio posteriors for the three planets and the transit depth posteriors from the \HST transits for Kepler 51b and 51d. For Kepler 51c's radius ratio, we used the value and uncertainties determined by \citet{masuda.2014}.\deleted{We created a Gaussian radius ratio posterior for Kepler 51c from the values cited in \citet{masuda.2014}.} The masses, radii, and densities for the Kepler 51 planets are shown in Table~\ref{tab:mr_params}. We verify that all three planets have extremely low-densities, less than 0.1 $g/cm^{3}$. We also found that the density of Kepler 51d is about $2\sigma$ smaller and Kepler 51b is $1\sigma$ greater than those values given by \citet{masuda.2014}. Knowledge of the correct densities of Kepler 51b and 51d is critical for the interpretation of their spectra presented in the next section.

\begin{table}[]
\centering
\caption{Planetary masses, radii, and densities for Kepler 51b, 51c, and 51d. Values were determined by combining the mass and radius for Kepler 51 with the mass ratios for each planet from the TTV analysis and radius ratios from the \HST transits for Kepler 51b and 51d and \kepler transits for Kepler 51c.}

\begin{tabular}{lccc}
                                            & Kepler 51b                   & Kepler 51c                   & Kepler 51d        \\ \hline
\multicolumn{1}{l|}{Mass ($M_{\oplus}$)}   & $3.69^{+1.86}_{-1.59}$    & $4.43\pm0.54$              & $5.70\pm1.12$   \\
\multicolumn{1}{l|}{Radius ($R_{\oplus}$)} & $6.89\pm0.14$              & $8.98\pm2.84$              & $9.46\pm0.16$   \\
\multicolumn{1}{l|}{Density ($g/cm^{3}$)}   & $0.064\pm0.024$ & $0.034^{+0.069}_{-0.019}$ & $0.038\pm0.006$
\end{tabular}
\label{tab:mr_params}
\end{table}

\section{Results}\label{sec:results}

\subsection{Transmission Spectra Results} \label{subsec:trans_results}

\begin{figure}[h]
\includegraphics[width=\textwidth]{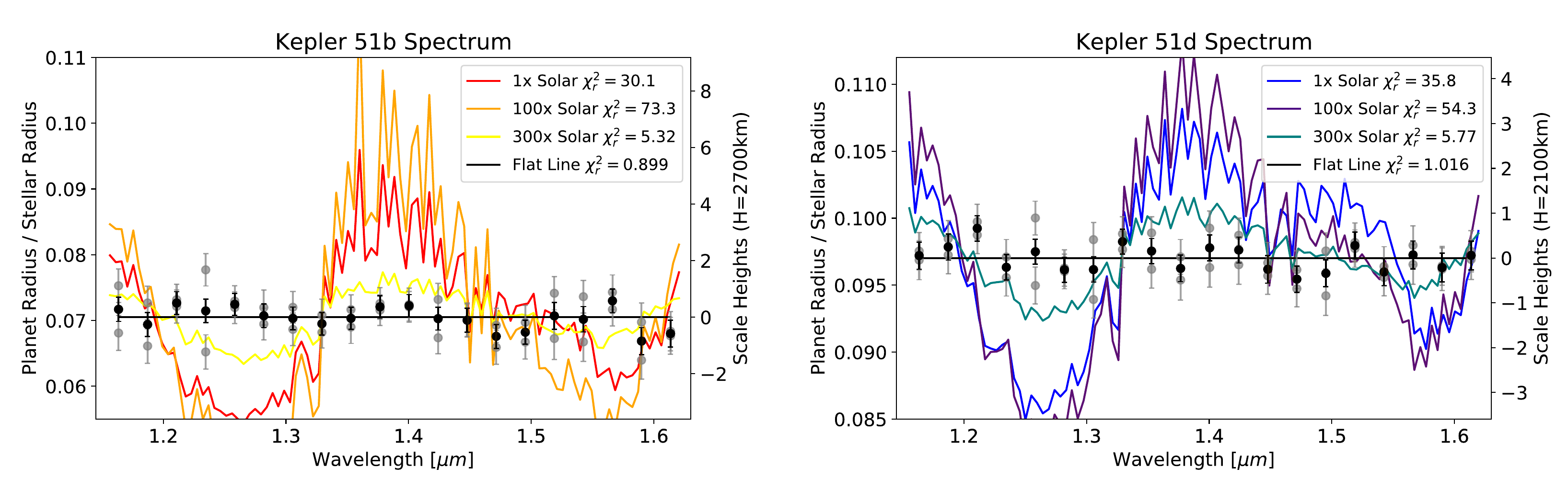}
\centering
\caption{The transmission spectrum for Kepler 51b on the left and Kepler 51d on the right from the WFC3 observations. Gray points represent the $R_{p}/R_{s}$ radius ratio determined for each independent visit. The weighted average between these two visits is plotted in black. Cloud-free models at varying metallicities are also plotted for comparison along with their corresponding $\chi^{2}_{r}$ with 19 degrees of freedom. The best model representing the data for both planets is a flat line with no spectral absorption features.}
\label{fig:kep51_spectra}
\end{figure}

We plot the WFC3-determined $R_{p}/R_{s}$ as a function of wavelength for Kepler 51b and Kepler 51d in Figure~\ref{fig:kep51_spectra}. We found that both planets' transmission spectra are flat, with a horizontal line yielding a $\chi^{2}_{r}$ of 0.90 and 1.02, respectively, with 19 degrees of freedom. From the weighted uncertainties, we approximated that the $1\sigma$ precision corresponds to 0.6 scale heights (1600~km) for Kepler 51b and 0.3 scale heights (630~km) for Kepler 51d. We compared our measurements to cloud-free transmission models of varying metallicities, ranging from 1-300x solar. Model pressure-temperature (P-T) profiles were calculated assuming radiative-convective and chemical equilibrium, as described  in \citet{McKay89, Marley96, Fortney08, Morley12, Morley15}. Our opacity database for gases is described in \citet{Freedman08, Freedman14}. We calculate the transmission spectrum for each converged P-T profile by calculating the optical depths for light along the slant path through the planet’s atmosphere at each wavelength, generating an effective planet radius at each wavelength. The model is described with more detail in \citet{fortney.2003} and \citet{shabram.et.al.2011}. The transmission spectrum model accounts for the fact that surface gravity varies with height in a planet, which is particularly important for these `super-puffs' where one scale height corresponds to roughly 3\% of each planet's radius.

We rule out cloud-free atmospheres with metallicities below 300x solar with greater than $7\sigma$ confidence for both Kepler 51b and 51d. Atmospheres with higher metallicities, and thus larger mean molecular weights and smaller scale heights, could potentially have small enough absorption features to fit the flat spectra. \added{We will explore this theory as well discuss alternative explanations for these transmission spectra in Section ~\ref{sec:discussion}.} 

\deleted{However, the 300x solar atmosphere model corresponds to an already large mean molecular weight of 6.8 amu. Given the large H/He contents required to explain the bulk densities of these planets, it is unlikely that atmospheres with mean molecular weights larger than this would be consistent with the masses and radii of Kepler 51b and 51d.}


The featureless spectra of both planets could also potentially be due to incorrect planetary masses which would lead to wrong densities and scale heights. Assuming a 1x solar atmospheric composition, we find that the mass for Kepler 51b would need to increase by more than $10\sigma$ over our reported value in order to shrink its modeled features to be consistent with the observed flat spectrum. For Kepler 51d, we would require the mass to be wrong by a factor of $17\sigma$. In our re-analysis of the masses, radii, and densities of all three planets (Section ~\ref{sec:mr}), we found no evidence for such catastrophic errors. Therefore, we dismiss this hypothesis; the featureless spectra are not due to erroneous planetary mass measurements.

\subsection{Comparison of \kepler and \HST Determined Planetary and Stellar Parameters}\label{sec:kepler_compare}

We compare the planetary parameters we derive from both \HST transits and our individual \kepler transit analysis (Section ~\ref{sec:kepler_analysis}) to the parameters listed in \citet{masuda.2014}. We observed no variation between the values of $a/R_{s}$ or inclination determined from the \HST transits and those from individual fits to each transit in the \kepler data set, for any of the Kepler 51 planets.

\begin{figure}[h]
\includegraphics[width=\columnwidth]{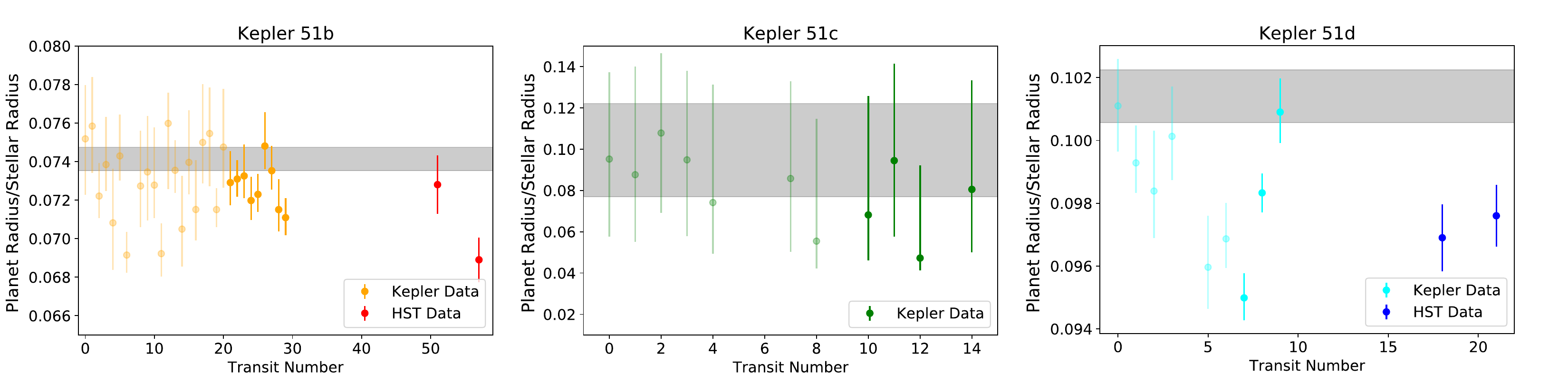}
\centering
\caption{The best-fit radius ratio ($R_{p}/R_{s}$) as a function of transit epoch for Kepler 51b, 51c, and 51d. The gray region represents the $1\sigma$ confidence limit reported in \citet{masuda.2014}. While \HST and \kepler observe different wavelength bandpasses, both instruments observed scatter between different transit epochs which we attributed to stellar activity. The large variations across all epochs makes it difficult to compare between the transit depths of each planet in the broadband optical \kepler wavelengths and \HST's infrared wavelengths. To search for a Rayleigh scattering slope in the atmospheres of Kepler 51b or 51d, a more precise simultaneous optical/IR measurement taking into account stellar variability would be required.}
\label{fig:kep51_radius_params}
\end{figure}

Figure~\ref{fig:kep51_radius_params} details the variation in the best-fit $R_{p}/R_{s}$ parameter over transit epoch, including both \kepler and \HST transits. While the radius measurements for Kepler 51c fall within the large $1\sigma$ error bars from \citet{masuda.2014}, Kepler 51b and 51d demonstrate a large amount of scatter in their respective radii over time. In particular, only 3 transits for Kepler 51d have an inferred $R_{p}/R_{s}$ within $1\sigma$ of the average value from \citet{masuda.2014}. Instead, the transits appear to fluctuate in radii with a difference of approximately 0.6\% $R_{p}/R_{s}$ between the largest and smallest measurements. This difference is equivalent to a change of approximately 0.05 $R_{\oplus}$ in the size of Kepler 51d. Kepler 51b also demonstrates scatter in its size versus transit epoch. However, most points fall within the \citet{masuda.2014} $1\sigma$ uncertainty limit. We attribute the temporal scatter in radii for Kepler 51b and 51d witnessed in Figure~\ref{fig:kep51_radius_params} as likely due to changing flux from stellar activity for each transit epoch. Assuming the same magnitude of stellar activity would lead to about 0.7\% $R_{p}/R_{s}$ discrepancy in the radius ratio for Kepler 51b, which is not observed in the \kepler light curves. The second observation from \HST does posses a radius ratio that varies by this amount, and we thus attribute this difference to stellar activity as well. With Kepler 51c's grazing geometry, we do not have the precision necessary to determine stellar activity of this magnitude.\deleted{This conclusion is also supported by both the number of star-spot crossing events observed in \kepler and \HST light curves as well as Kepler 51's young age.}

Because of this activity, we recommend caution when comparing the transit depth of Kepler 51b and 51d observed in the \kepler optical bandpass to those observed in the \HST WFC3 bandpass. On first glance, it may appear that Kepler 51d posseses a smaller radius in the \HST observations compared to the value reported in \citet{masuda.2014}. However, this smaller radius is still reproduced in some of the individual \kepler transits, and at times, the \kepler light curves demonstrate an even smaller radii for Kepler 51d than what we observed in either of the \HST visits. The same argument can be made for Kepler 51b, with one of the \HST visits falling well within the $1\sigma$ uncertainty limit, while the other does not. In the face of this stellar activity, it is challenging to measure a Rayleigh scattering slope between the wavelengths of WFC3 (1.4~$\mu$m) and \kepler (0.6~$\mu$m), because the transits were not observed simultaneously and the impact of starspots on the transit depths is unknown \citep{rackham.et.al.2019}. Still, we place quantitative upper limits on any hypothetical scattering slope for each planet by averaging the individual \kepler light curve transit depths and comparing this value to an average of the \HST visits. To account for the non-simultaneity, we adopt the standard deviation of the individual transit depth measurements as the errors on these two wavelength-averaged transit depths. We determined slopes ($\delta/\ln{\lambda}$) of $-0.0004\pm0.0004$ and $-0.0002\pm0.0011$ for Kepler 51b and 51d respectively based on these two points. If a scattering slope is present below our cautious error bars, more careful observations and/or treatment stellar activity would be required to make any confident conclusion regarding the existence of such a slope for either planet.

\begin{figure}[h]
\includegraphics[width=\columnwidth]{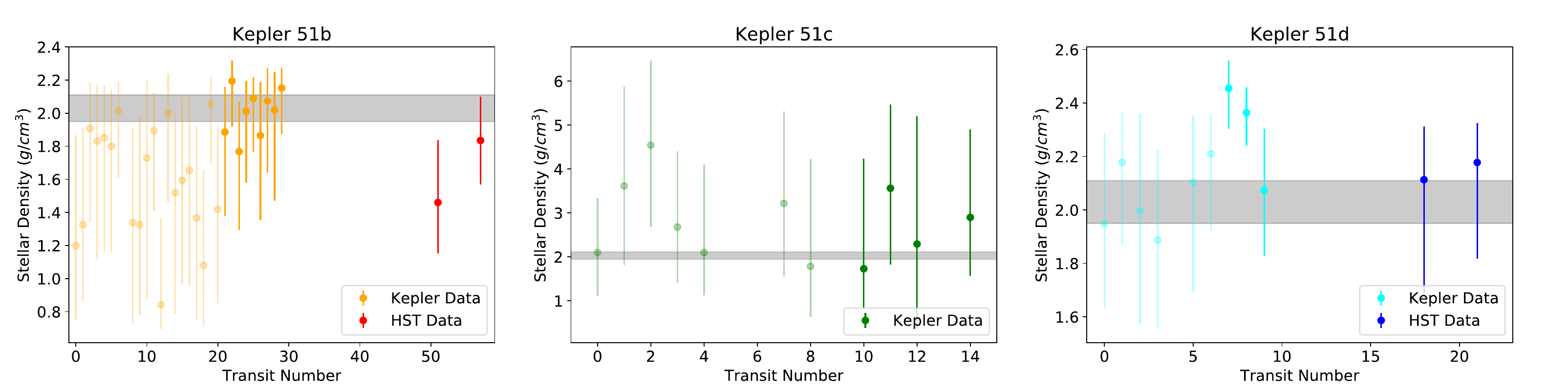}
\centering
\caption{Stellar densities from the individual transit light curve fits for each planet, compared to the $2.03\pm0.08 g/cm^{3}$ independently-estimated stellar density from {\tt isochrones}. Lighter points represent \kepler's long cadence data, while darker points of the same color represent short cadence \kepler data. We include the measured eccentricity for all three planets in the stellar density calculations using the correction factor in \citet{dawson.et.al.2012}. Most \kepler transits and all of the \HST transits support the stellar density (agree within 2$\sigma$) determined from {\tt isochrones}.}
\label{fig:kep51_stellar_params}
\end{figure}

\added{We use both the \kepler and \HST transits for all three Kepler 51 planets to calculate the stellar density by combining th $a/R_{s}$ posterior at each transit epoch with the posterior for each planet's period determined from our TTV analysis. We corrected for eccentricity when determining the stellar density by multiplying the circular stellar density values by the appropriate factor of $((1+e\sin\omega)/\sqrt{(1-e^{2})})^{-3}$, as shown in \citet{dawson.et.al.2012}. A comparison of these stellar density values to the stellar density determined by combining the stellar mass and radius posteriors from {\tt isochrones} demonstrates good agreement between the two methods (Figure ~\ref{fig:kep51_stellar_params}). Comparing the {\tt isochrones} density to that reported in \citet{masuda.2014} also shows less than a 1$\sigma$ discrepancy in the two values. The consistency of all available methods for estimating stellar density leads us to conclude that Kepler 51 is a single, young, Sun-like, main-sequence star.}

\section{Discussion}\label{sec:discussion}

\subsection{Pure H/He Atmosphere}

\begin{figure}[h]
\includegraphics[width=\textwidth]{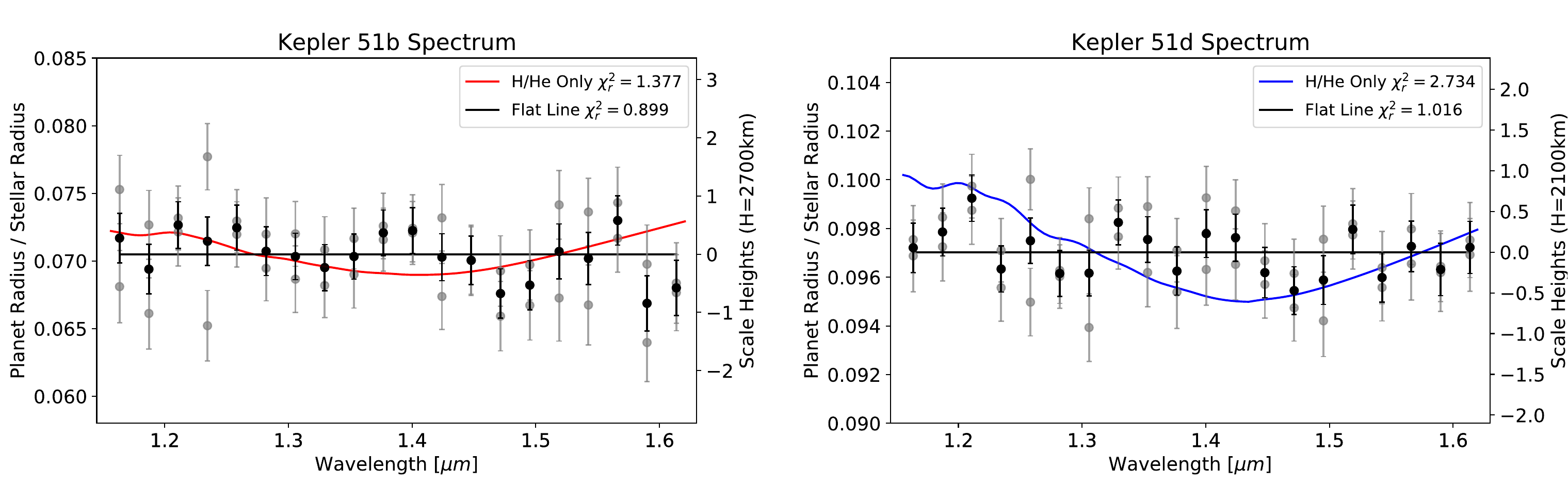}
\centering
\caption{The transmission spectra for Kepler 51b (left) and Kepler 51d (right), compared to models of pure H/He atmospheres. Gray points represent the $R_{p}/R_{s}$ radius ratio determined for each independent visit. The weighted average between these two visits is plotted in black. Features in the model are due to Rayleigh scattering and collision-induced absorption from the hydrogen and helium molecules in the atmosphere; all other absorption features were removed. We confidently rule out a H/He-only model for Kepler 51d but not for Kepler 51b.}
\label{fig:kep51_hhe_spectra}
\end{figure}

Given their bulk densities alone, Kepler 51b and 51d could potentially possess a metal-free, pure H/He atmosphere. We test this hypothesis by creating a H/He-only model to compare to the spectra of both planets. To do this, we use the 1x solar composition model for each planet and remove absorption features from any gases outside of molecular hydrogen and helium, leaving only Rayleigh scattering and collision-induced absorption features. Figure ~\ref{fig:kep51_hhe_spectra} compares the transmission spectra to these H/He models for both Kepler 51b and 51d. We are able to confidently dismiss a strictly H/He atmosphere for Kepler 51d with a $\chi^{2}_{r}$ of 2.7 which corresponds to a $p$-value of $< 0.001$. We cannot rule out a H/He-only atmosphere for Kepler 51b ($\chi^{2}_{r}$=1.4, a $p$-value of 0.1). However, as the formation of completely metal-free atmospheres has no known theoretical basis, we do not find a pure H/He atmosphere to be the most plausible explanation of Kepler 51b's flat spectrum.

\subsection{Clouds and Hazes}
\label{sub:clouds}

\begin{figure}[h]
\includegraphics[width=\columnwidth]{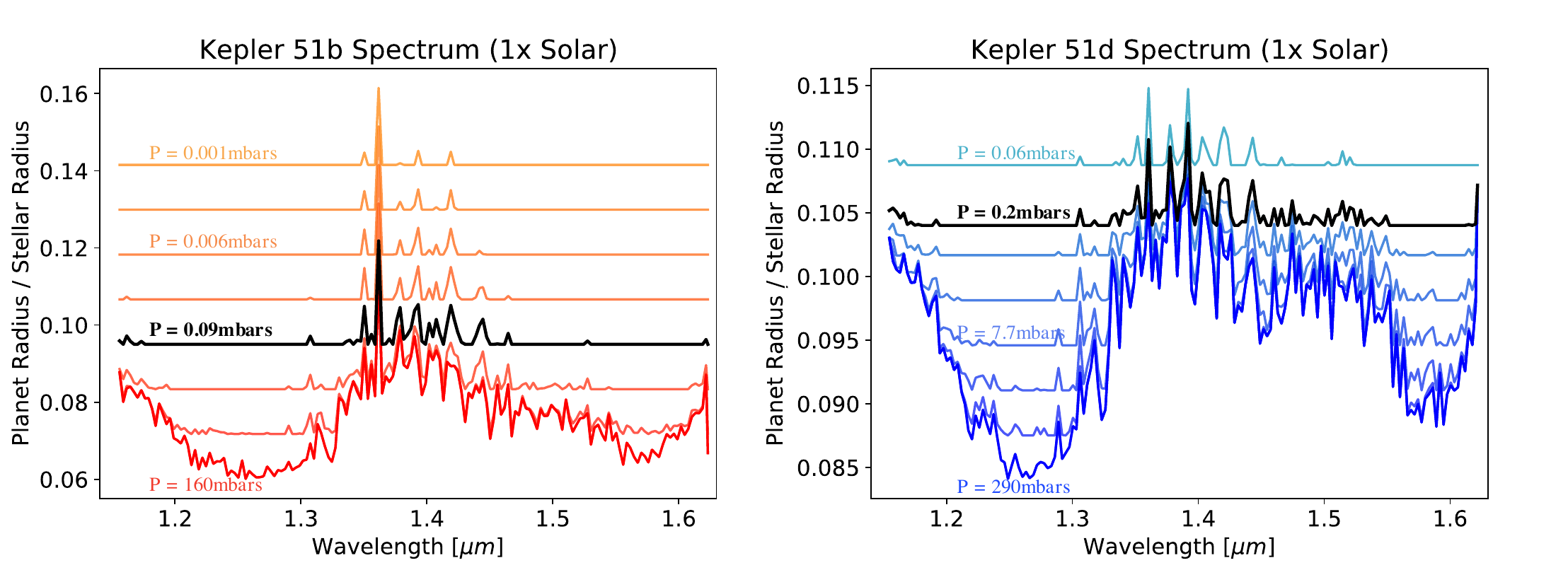}
\centering
\caption{An example of aerosol models at varying altitudes and pressures for both Kepler 51b and 51d. We continue to increase the altitude (and decrease the pressure) for this aerosol layer until the corresponding model matched the data within a 95\% confidence limit (black line). Any layer above this pressure level will continue to match the flat spectra. Given the flatness of the spectra, both Kepler 51b and 51d must have an aerosol layer at pressures of less than 0.09 and 0.2 mbars respectively assuming a 1x Solar atmospheric metallicity. These pressure levels decrease to microbars assuming 100x Solar metallicity.}
\label{fig:cloud}
\end{figure}

We propose that both Kepler 51b and 51d maintain an opaque high-altitude aerosol layer in their atmospheres which is flattening their transmission spectra in the WFC3 bandpass. We explore this theory by taking a cloud-free atmospheric model and subsequently adding an aerosol layer at higher and higher altitudes until we determined that the flattened modeled spectrum matches the data within a 95\% confidence limit (Figure~\ref{fig:cloud}). We assumed this cloud layer to be composed of an unknown gray absorber which mutes any absorption features falling below it while not contributing any of its own absorption or emission features. We repeat this exercise for both Kepler 51b and 51d models assuming metallicites of 1x, 100x, and 300x solar. Table~\ref{tab:cloud_alt} highlights the pressure limits required for an aerosol layer at varying atmospheric metallicities. We emphasize that these are upper limits on the pressure levels, as any aerosol layer at lower pressures than these (higher altitudes) will also flatten the spectrum for each planet.

\begin{table}[]
\centering
\begin{tabular}{lc|c}
\multicolumn{1}{c}{}             & Kepler 51b & Kepler 51d \\ \hline
\multicolumn{1}{c|}{Metallicity} & P (mbars)  & P (mbars)  \\ \hline
\multicolumn{1}{l|}{1x Solar}    & 0.09       & 0.2        \\
\multicolumn{1}{l|}{100x Solar}  & 0.001      & 0.002      \\
\multicolumn{1}{l|}{300x Solar}  & 0.1        & 0.09      
\end{tabular}
\caption{Upper limits (95\% confidence) on the pressure level for the aerosol layer at the corresponding atmospheric metallicities for both planets.}
\label{tab:cloud_alt}
\end{table}

We find that 1x and 300x solar metallicities required an aerosol layer at pressure levels around 0.1 millibars. However, if the atmospheres are 100x solar, the aerosol layer must occur at pressure levels of less than 2 microbars for both planets. This change in aerosol altitude between 1x and 100x solar is due to the increased strength in the molecular absorption features while the mean molecular weight of the atmosphere remains largely unchanged. \citet{crossfield.and.kreidberg.2017} note that once an atmosphere becomes greater than 100x solar, the mean molecular weight will increase rapidly, diminishing the amplitude of the features.

\begin{figure}[h]
\includegraphics[width=8cm]{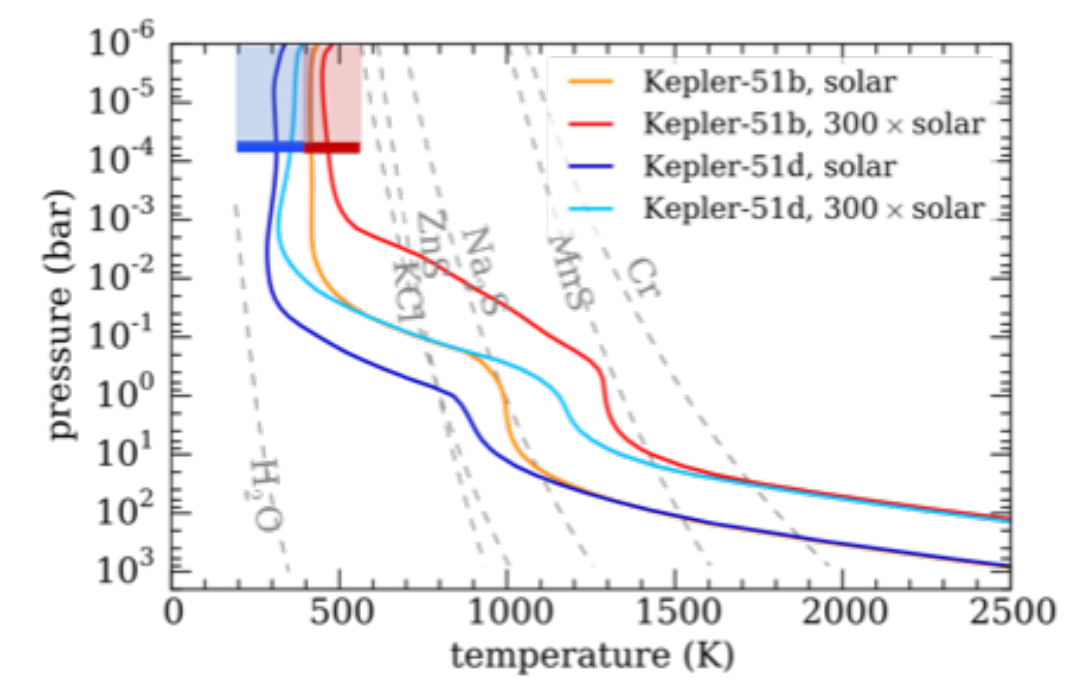}
\centering
\caption{The pressure-temperature profile for Kepler 51b and 51d at varying metallicities. Shaded in red for Kepler 51b and blue for Kepler 51d are the pressure levels above which aerosols must be present to match the observe WFC3 transmission spectra. Neither profile crosses a condensation curve at the required pressure levels, and we are therefore unable to determine the aerosol composition from this alone. Based on their high altitude, we argue photochemical hazes are a more likely source for the aerosols we see blocking transmission.}
\label{fig:pt_profile}
\end{figure}

We compared the determined pressure levels to simulated pressure-temperature profiles for Kepler 51b and 51d (Figure~\ref{fig:pt_profile}). We would expect the condensation of cloud particles from abundant molecules KCl, ZnS, and Na$_2$S to form near the altitudes where the planets' P-T profiles cross the condensation curves for these species \citep{sanchez-lavega.2004.cpauace}. For both planets, the P-T profiles cross these condensation curves, but at much deeper altitudes than where we require an aerosol layer. If condensates are responsible for the aerosols we see, strong vertical mixing would be required to loft particles from where they are created up to higher altitudes. \added{Combining models from \citet{Morley15} with the large scale heights of Kepler 51b and 51d, we note that a vertical eddy diffusion coefficient greater than $10^{10}$ cm$^{2}$/s is required for lofting 1 $\mu m$ sized particles up to pressures of 0.1 mbars. For comparison, \citet{dong.et.al.2016} predict vertical eddy diffusion coefficients between $10^{7}$ and $10^{9}$ cm$^{2}$/s for Jupiter and Saturn.}

However, given the relatively cool equilibrium temperatures of 550~K and 360~K for Kepler 51b and 51d respectively, it is likely that the main carbon-containing molecule is methane \citep{heng.2017}. At a young age, it is likely that Kepler 51 is outputting a large amount of XUV flux \citep{xuv.paper}. The combination of methane and these UV photons may be producing a substantial photochemical haze layer on both planets, similar to that observed on Titan \citep{horst.2017}. Furthermore, \citet{heng.2017} discusses that if \replaced{$C_{2}H_{2}$}{{$\rm C_{2}H_{2}$}} is present in the atmosphere at pressures of 10 milibars or less (as is the case with Jupiter), planets with equilibrium temperatures around 500~K will experience hydrogenation leading to the creation of large amounts of higher-order hydrocarbons such as \replaced{$C_{2}H_{4}$}{$\rm C_{2}H_{4}$} and beyond. In turn, these hydrocarbons could create a photochemical haze layer \citep[e.g.][]{morley.et.al.2013}. \citet{kawashima.et.al.2019} found that haze particles produced from methane photodissociation on Kepler 51b could grow to sizes of $~$0.1~$\mu$m at pressure levels of 1 microbar up to 100~$\mu$m at 10 bars. They attribute this large growth to the low surface gravities which allows large particles to remain at high altitudes. Due to these particle sizes, their haze models predict a shallower Rayleigh slope than a typical hazy atmosphere as well as potentially muted \added{absorption features of water, carbon monoxide, and carbon dioxide absorption features at longer wavelengths (beyond 2 $\mu$m) that may be accessible by the James Webb Space Telescope.} \deleted{but not completely featureless transmission spectrum out to large wavelengths.} 

Similarly, \citet{gao.et.al.2017} suggest that Jupiter-mass planets with temperatures between 250~K and 700~K may be subjected to photochemical dissociation of \replaced{$H_{2}S$}{$\rm H_{2}S$}, which could produce a haze layer of \replaced{$S_{8}$}{$\rm S_{8}$} somewhere between 100 and 0.1 millibars. Determining the composition of the aerosol layers on Kepler 51b and 51d could provide an important step towards understanding which processes dominate in the atmospheres of warm sub-Neptunes. For now, we argue that due to the combination of the low surface gravities of the planets, the probable presence of methane in their atmospheres, and the isothermal profile at the low pressures required for an aerosol layer, we are likely dealing with a photochemical haze layer and not a cloudy condensation layer. 

\subsection{Alternative Hypotheses}
\label{sub:alt}

A \added{large atmospheric mean molecular weight can create a flattened spectrum by shrinking the scale height and thus the amplitude of the features \citep{millerricci.et.al.2009}. To test this hypothesis, we determined the maximum atmospheric mean molecular weight possible given Kepler 51b and 51d's low-densities. Assuming a 1 $M_{\oplus}$ core for both planets while the rest of the mass is well-mixed in the envelope, we calculated a maximum mean atmospheric molecular weight of 27 amu and 21 amu for Kepler 51b and 51d respectively. This maximum weight is sensitive to the size of the core of each planet, which is not well-constrained given current formation theories. A larger core would lead to smaller possible mean molecular weights. The 300x solar atmosphere model plotted in Figure ~\ref{fig:kep51_spectra} corresponds to an atmosphere with 7 amu. However, we find that we are unable to confidently rule out the maximum mean atmospheric weight for both planets as the cause for the flat spectra. More precise observations with future instruments will be required to compare the two hypotheses: aerosol layer or a high mean molecular weight atmosphere.}

\citet{wang.and.dai.2019} discuss another possible mechanism for flattening a spectrum. Planets with low surface gravities that are currently undergoing hydrodynamic mass loss may also be experiencing outflows of dust particles. These dust particles, with sizes less than 10 angstroms, would both mask absorption features and create an overestimate of the actual planetary radius. By modeling a Kepler 51b-like planet, \citet{wang.and.dai.2019} found that their dusty outflow hypothesis is testable by observing the pre-transit baseline and ingress of Kepler 51b with a precision of 200~ppm. If dust is present, they predict the start of the ingress to demonstrate a gentle slope compared to a typical transit. Future telescopes, notably \replaced{JWST}{the James Webb Space Telescope (JWST)}, might reach the precision required to distinguish between our haze-layer hypothesis and the dusty-outflow model of \citet{wang.and.dai.2019}.

\subsection{Evolution of the Mass and Radius of Kepler 51b and 51d}
\label{sec:evolution}

\begin{figure}[h]
\includegraphics[width=15cm]{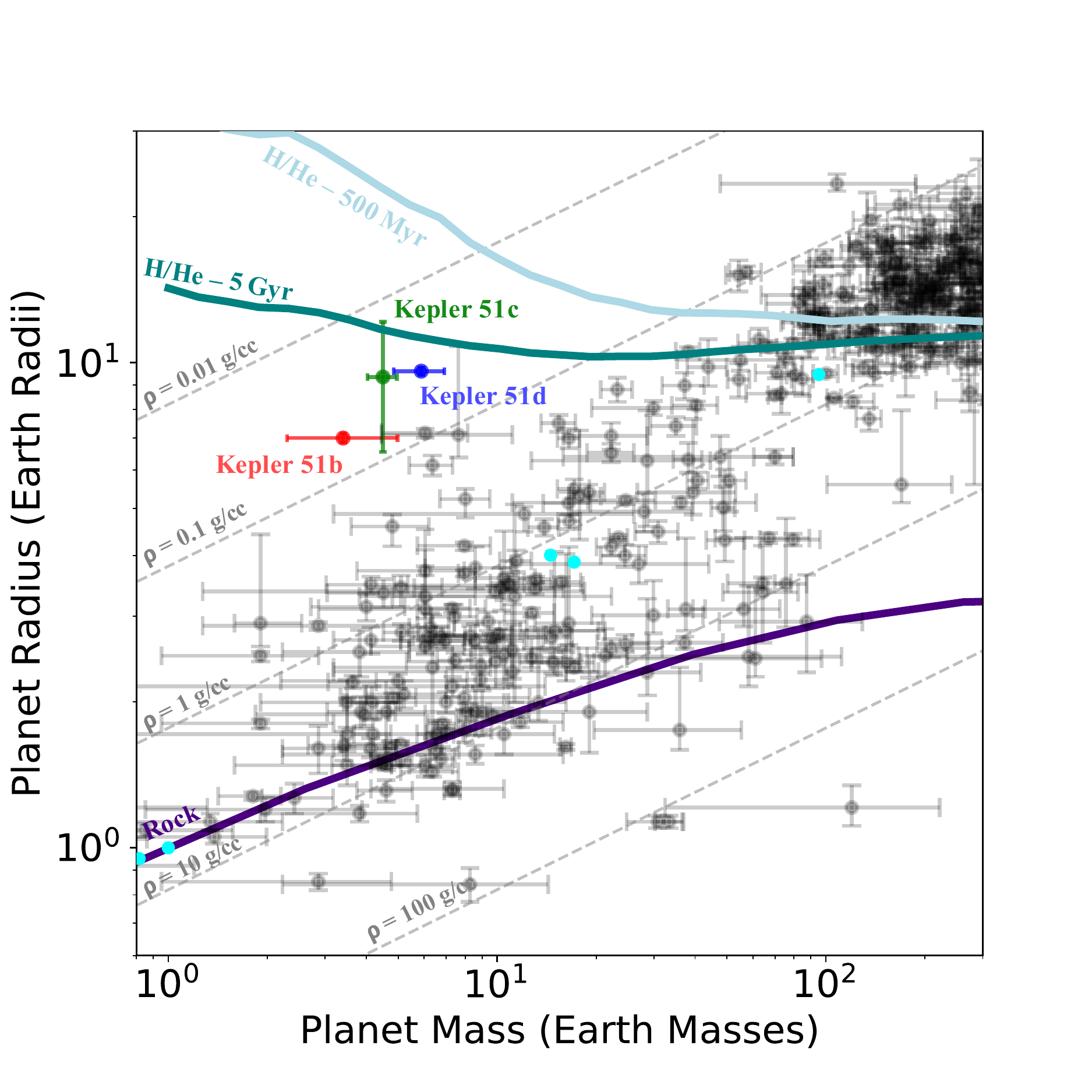}
\centering
\caption{Mass/radius diagram of known exoplanets. Kepler 51b, 51c, and 51d are plotted with red, green, and blue points respectively. Solar system planets are included as bright blue points for reference. Lines of constant density are plotted in gray and labeled on the left of the plot. We also include the mass-radius model for a rocky composition from \citet{seager.2007} (purple) and a pure 100\% H/He composition for a 500 Myr (light blue) and 5 Gyr (teal) old planet based on modified models presented in \citet{lopez.and.fortney.2014}. For the H/He models, we assumed an insolation similar to Kepler 51b, while the rocky model is independent of insolation. All three Kepler 51 planets fall in a region of the mass-radius diagram sparsely populated by other planets. \replaced{With their mass-radius combinations, these planets fall above the cold H/He line. However,the Kepler 51 planets are still evolving towards their final masses and radii.}{However, age affects the masses and radii of planets as demonstrated by the discrepancy between the two H/He curves. We therefore predict that the Kepler 51 planets are still evolving to their final mass/radius.}}
\label{fig:mr_plot}
\end{figure}

\deleted{Figure~\ref{fig:mr_plot} compares the masses and radii of the Kepler 51 planets to other planets with known masses and radii, highlighting their extreme low-densities. All three Kepler 51 planets fall in the top left portion of Figure \ref{fig:mr_plot}: a sparsely populated region well above the cold H/He gas mass-radius relationship model calculated by \citet{seager.2007}. As gas planets cool over time, they will also contract. The Kepler 51 planets are relatively young and inflated, but they will likely eventually move below the cold H/He density line in Figure \ref{fig:mr_plot} due to their contracting envelopes \citep{rogers.et.al.2011,lopez.and.fortney.2014}.}

\added{Figure~\ref{fig:mr_plot} compares the masses and radii of the Kepler 51 planets to other planets with known masses and radii, highlighting their extreme low-densities. All three Kepler 51 planets fall in the sparsely populated region on the top left of Figure~\ref{fig:mr_plot}. However, both planets fall below the pure H/He gas mass-radius relationship models for both a 500 Myr and 5 Gyr old planet with the insolation of Kepler 51b \citep{lopez.and.fortney.2014}. Even with their extreme densities, it is physically plausible from a structural standpoint for both of these planets to exist. In fact, as these planets fall below the H/He curves at both ages, they must posses a higher density composition whether it is in the core, envelope, or a combination of both.}

\begin{figure}[h]
\includegraphics[width=15cm]{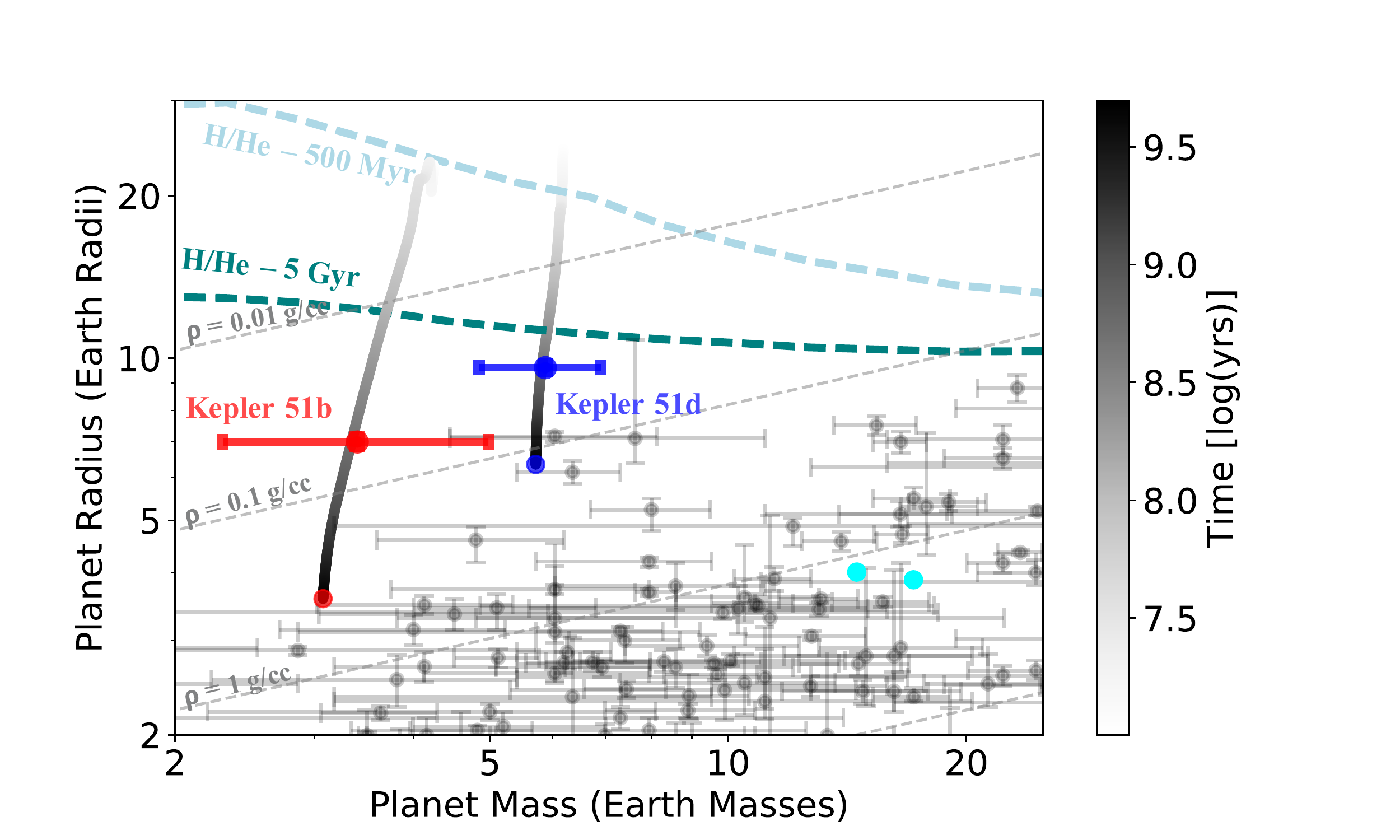}
\centering
\caption{Predicted evolution of the masses and radii of Kepler 51b and 51d from 10 Myr to 5 Gyr, based on the \added{modified} models of \citet{lopez.and.fortney.2014}. The 5 Gyr masses/radii for both planets are plotted as a red point for Kepler 51b and a blue point for Kepler 51d at the end of the evolution curve. Current masses and radii for the two planets are also included with error bars at the 500 Myr mark on each curve. We again include lines of constant densities and \replaced{the \citet{seager.2007} mass-radius prediction for a H/He-dominated planet}{the mass-radius for a 100\% H/He planet at 500 Myr and 5 Gyr from Figure ~\ref{fig:mr_plot}. As Kepler 51b and 51d likely have a core of some mass, these H/He models are provided as reference, not an explanation of the possible composition of the low-density Kepler 51 planets}. Other exoplanets with known masses and sizes are also included as gray points while Uranus and Neptune are noted by the two cyan points. Note that the color of these points does not correspond to the time color bar on the right. By 5 Gyr, Kepler 51b will lose a substantial amount of mass while also contracting to the size of a sub-Neptune sized planet. Kepler 51d will retain most of its mass, though it too will contract significantly, still possessing a density that is lower than typical but similar to other super-puffs like Kepler 79d. \added{We do not include the substantial uncertainties in the evolutionary models for each planet, but the qualitative conclusion is clear: these planets will increase in density over time due to ongoing contraction and mass loss, approaching the larger population of other known low-mass, low-density planets.}}
\label{fig:evolution}
\end{figure}

Most planets plotted in Figure~\ref{fig:mr_plot} are much older than the Kepler 51 planets. \added{However, age of a system can significantly shape the masses and radii of a planet especially at smaller masses (Figure ~\ref{fig:mr_plot}).} This \replaced{discrepancy}{discrepancy in planet ages when comparing planetary populations} leads us to the question: given the young 500 Myr age of the system, are the extremely low densities of the Kepler 51 planets really that unusual? \replaced{Using}{Modifying the} thermal evolution and photoevaporation models presented in \citet{lopez.and.fortney.2014} to accommodate for the low surface gravities, we modeled the the evolution of the masses and radii of planets similar to Kepler 51b and 51d from 10 Myr to 5 Gyr \added{for a 1x solar metallicity atmosphere} (Figure~\ref{fig:evolution}). For these models, we assumed a young Sun-like star output which also evolves over the 10 Myr to 5 Gyr time period. We find that Kepler 51b will lose a significant amount of mass at a current rate of $10^{10}-10^{11}$ g/s. \added{However, multiple uncertainties in these models are not quantifiable accounted for: the largest of which is the evolution of the planet masses and the XUV flux of the star over time. As both play an important role in mass-loss rates, we expect roughly an order of magnitude uncertainty on the mass loss rates quoted above.} With that caveat, our models predict that Kepler 51b will start with H/He with a mass fraction of 36\%, but with photoevaporation, it will possess a H/He mass fraction of only 11\% at 5 Gyr. Kepler 51d will retain most of its H/He starting with a mass fraction of 39\% and ending up with 34\%. However, both planets will shrink considerably over this time with Kepler 51b ending up at 3.7 $R_{\oplus}$ and 51d having a radius of 6.2 $R_{\oplus}$. With the combined mass loss and contraction, we predict that Kepler 51b will look akin to many other sub-Neptune planets at 5 Gyr. Kepler 51d, on the other hand, will remain a low-density planet, but still similar to other low-density planets such as Kepler 79d \citep{jontof-hutter.et.al.2014}. 

\begin{figure}[h]
\includegraphics[width=15cm]{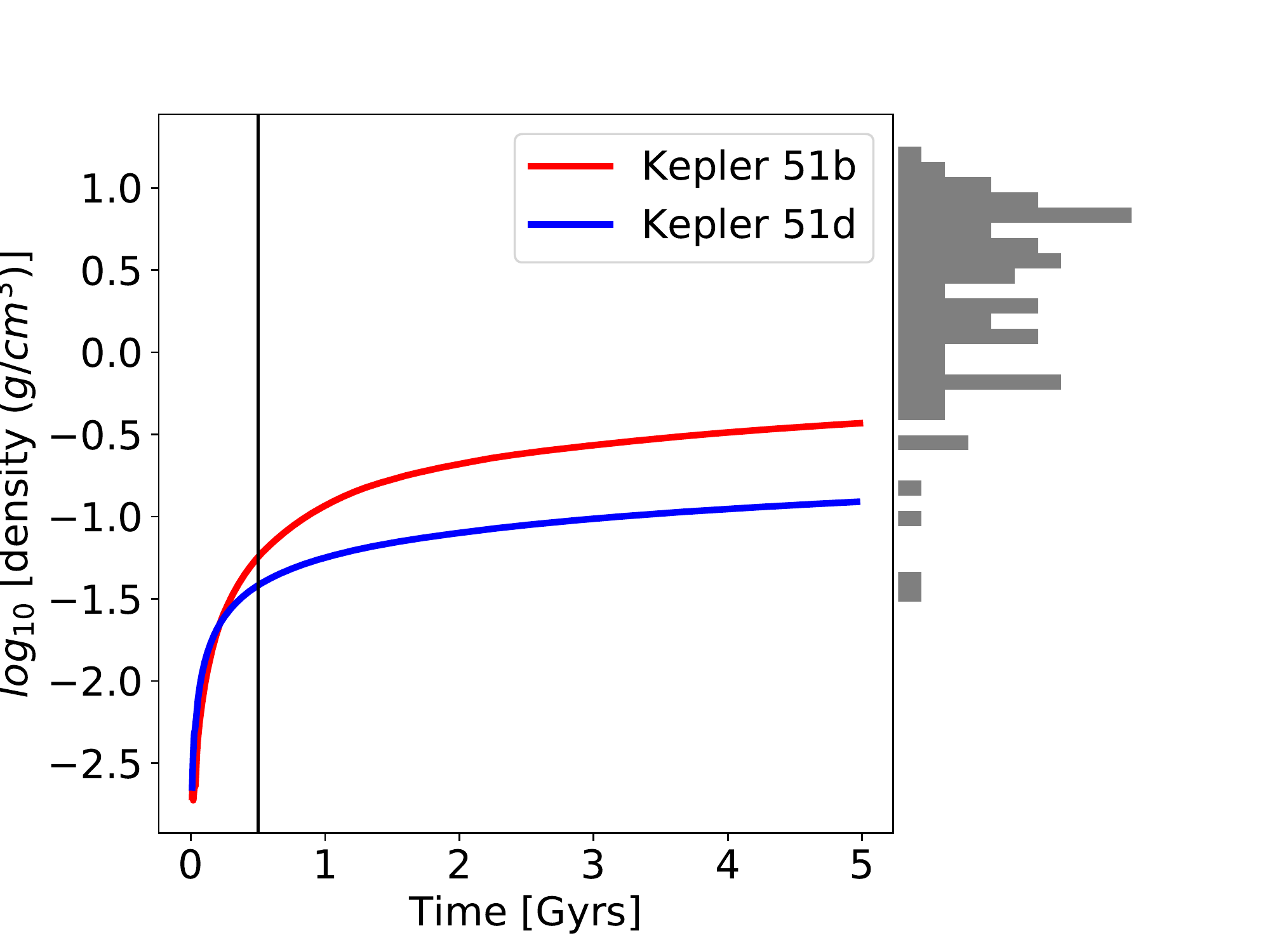}
\centering
\caption{The predicted density of Kepler 51b (red) and 51d (blue) as a function of time determined from the same evolution curves shown in Figure~\ref{fig:evolution}. The vertical black line marks the current 500 Myr age of the system. A histogram of all exoplanets with known densities and masses less than 20~$M_{\oplus}$ is plotted to the right. As Kepler 51b and 51d undergo contraction and mass-loss, they will eventually possess densities similar to other exoplanets of similar masses, albeit on the lower end of the histogram.}
\label{fig:density_evolution}
\end{figure}

Figure~\ref{fig:density_evolution} plots the evolution curves shown in Figure~\ref{fig:evolution} for Kepler 51b (red) and 51d (blue) as density versus time. A vertical black line marks the current age of 500 Myr for the system. To the right of these curves is a histogram of the densities for all transiting exoplanets with known masses less than 20~$M_{\oplus}$. Based on the evolution curves, the densities of both Kepler 51b and 51d will slowly increase over the next 5 Gyrs due to both contraction and mass-loss. After 5 Gyr, both planets will reach densities comparable to other field-age low-mass planets, although still on the low end of the distribution. Even though Kepler 51b will undergo significant mass-loss, it will not lose its atmosphere completely and leave an evaporated core. In the end, both planets will likely evolve into low-density sub-Neptunes. 

Due to the evolution of masses and radii of the Kepler 51 planets, we argue that the formation of this system does not require substantial deviation from previous planet formation theories. Instead, the young age of the system is the driving feature of the large radii observed for all three planets. 

\added{All three of the Kepler 51 planets could currently be classified as low-mass, low-density, and low-irradiation super-puffs. However, as density is intrinisically linked to the age of a planet, a planet such as Kepler 51b, may eventually transition out of this population. The reverse could be stated: there may currently be sub-Neptunes observed that at one point started as low-density super-puffs. Even the density of Kepler 51d will increase due to planet contraction, giving it a similar density when compared to other low-density, low-irradiation, and older exoplanets. Therefore, it is possible that the three low-density planets of Kepler 51 are not as unusual as previously thought.} 

\deleted{We thus propose a modification to the \citet{lee.and.chiang.2016} definition of a super-puff, which was based on a planet's density. We consider any low-mass planet with a high mass fraction of H/He as a super-puff. Due to thermal contraction, density evolves with age so a super-puff might not retain its extreme low-density through time, even if its overall composition is relatively unchanged. Although a planet's H/He mass fraction may still decrease due to mass loss, it is more of an intrinsic property linked to formation than is density, particularly for planets like Kepler 51d that lose little of their mass over their lifetimes.}

\subsection{WFC3 Flat Transmission Spectra Comparison}

\citet{crossfield.and.kreidberg.2017} argue that the size of the water absorption feature of an exoplanet in the WFC3 bandpass may be related to either the equilibrium temperature of the planet or to its H/He mass fraction. They determined this possible relationship by analyzing six exoplanets with available WFC3 spectra. To compare these different sub-Neptunes, \citet{crossfield.and.kreidberg.2017} normalized each exoplanet's spectrum by the size of its scale height (assuming a H/He-dominate mean molecular weight). After a thorough analysis of multiple different parameters, only equilibrium temperature and H/He mass fraction demonstrated any statistically significant correlation.

Their first hypothesis is that hotter planets appear to have larger observable features, whereas cooler planets are able to more efficiently form high-altitude aerosol layers which mute absorption or block these absorption features. This relationship may indicate a link between aerosol formation and equilibrium temperature, as predicted by theoretical models \citep{fortney.et.al.2013,Morley15}. Their second hypothesis is based on the connection between exoplanets with a smaller H/He mass fraction having a high mean molecular weight atmosphere and therefore smaller amplitude features.

To test their two theories, \replaced{we determined the water absorption amplitude (or lack of one)}{we calculated an upper limit on the water absorption amplitude} for Kepler 51b and 51d using the same atmospheric model as \citet{crossfield.and.kreidberg.2017} and scaling the water feature amplitude until it matched the data by minimizing a $\chi^{2}_{r}$. \added{We determined a water amplitude feature of $-0.029 \pm 0.266$ and $-0.019 \pm 0.351$ for Kepler 51b and 51d respectively. We allowed our water amplitude to be negative in our fits as did \citet{crossfield.and.kreidberg.2017}.}

The H/He mass fraction of both planets came from models detailed in \citet{lopez.and.fortney.2014}:\added{ $17.5^{+5.2}_{-3.1}$\% and $35.2^{+7.6}_{-3.9}$\% for Kepler 51b and 51d respectively.} We calculated the equilibrium temperatures from the updated stellar parameters, assuming zero planetary albedo.

\begin{figure}[h]
\includegraphics[width=\columnwidth]{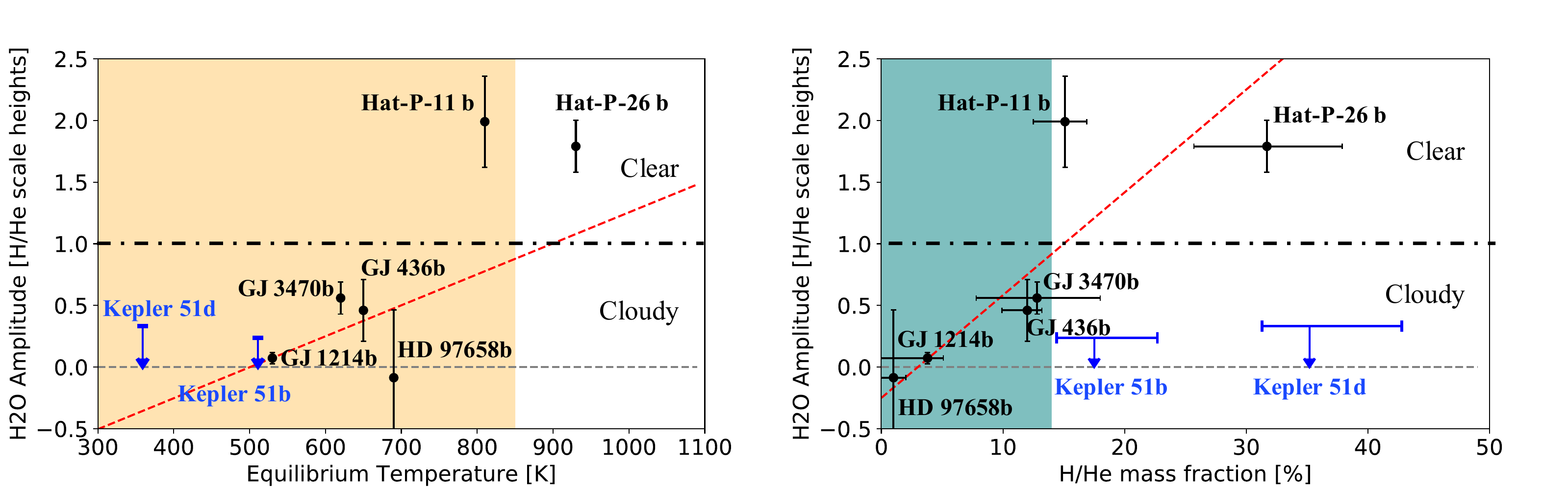}
\centering
\caption{Modified figure from \citet{crossfield.and.kreidberg.2017} highlighting two linear trends: a relationship between water amplitude features and equilibrium temperature (left), and a connection between water absorption amplitude and H/He mass fraction (right). Their six planets are plotted in black, while the Kepler 51b and 51d upper limits are plotted in blue. From the results, the Kepler 51 scenario does not support a first-order correlation between the amplitude of the absorption feature and the H/He mass fraction. However, the results do continue to support a link between temperature and muted absorption features due to aerosol formation.}
\label{fig:flat_spectra}
\end{figure}

We illustrate the new estimates for the Kepler 51 planets in Figure~\ref{fig:flat_spectra}, a modified version of Figures 2 and 3 in \citet{crossfield.and.kreidberg.2017}. As the only planets on these plots that are both H/He-rich and relatively cool, Kepler 51b and 51d provided important new tests for the \citet{crossfield.and.kreidberg.2017} relationships. The flat spectra of both Kepler 51b and 51d continue to support the scenario that cooler planets show weaker features. The exact shape of this relationship is still unknown; absorption feature strength may grow continuously across this temperature range, or it might increase sharply at equilibrium temperatures around 800~K. However, the Kepler 51 planets strongly refutes the scenario that a high H/He mass fraction directly guarantees a large water absorption feature in the observable WFC3 spectrum.

This analysis reinforces the picture that aerosol formation is somehow linked to equilibrium temperature, across the 400-1000~K temperature range. However, with the small sample size, the messy selection biases that decide which targets are observable with WFC3, and the fact that planetary atmospheres are complicated, using temperature as a single proxy to predict a signal size is likely ill-advised. There may still be other hidden variables controlling the appearance or absence of aerosols. 

\section{Conclusion and Summary}\label{sec:conclusion}

Kepler 51 is host to three of the lowest density exoplanets known to date. With densities less than $0.1g/cm^{3}$, these planets have atmospheric scale heights 10x larger than a typical hot-Jupiter making them prime targets for transmission spectroscopy. We observed the innermost (Kepler 51b) and outermost (Kepler 51d) transiting planets in this system with the \HST WFC3 G141 grism. The transmission spectra of both planets were flat, to the precision of these data. After updating the stellar and planetary parameters with the \HST observations and other new measurements, we independently verified the extremely low-densities of all three exoplanets originally quoted in \citet{masuda.2014}. We concluded that the flat transmission spectra must be due to high photochemical haze layer in the atmosphere of both planets.

We modeled the evolution of the masses and radii of Kepler 51b and 51d from 10 Myr to 5 Gyr using a photoevaporation and core contraction model \citep{lopez.and.fortney.2014}. We found that these planets, while currently possessing unusual densities, will both contract and lose some atmospheric mass over time. In the end, Kepler 51b will evolve to a fairly typical sub-Neptune while Kepler 51d will remain with a slightly lower than normal density. 

We compared the flat transmission spectra to the statistical analysis performed by \citet{crossfield.and.kreidberg.2017}. Both Kepler 51 planets fit their trend that cooler planets have more muted water absorption amplitudes (across equilibrium temperatures of 400-1000~K), likely due to more efficient aerosol formation in cooler atmospheres. As the first cool, H/He-rich planets added to the population, the Kepler 51 planets break with their trend that a higher H/He content will necessarily result in a spectrum with absorption features spanning more atmospheric scale heights. Future observations of other sub-Neptunes will allow for more in-depth statistical studies of aerosol formation in exoplanets, involving other crucial environmental inputs such as XUV irradiation and planetary densities/gravities.

While this paper presents the first transmission spectra of Kepler 51b and 51d, more work remains. The Kepler 51 planets provide a rare opportunity to study an exoplanet system in its teenage years where the system is formed but still undergoing significant evolution. Kepler 51 is also a fairly active young star, and we expect the probable excess EUV and X-ray flux to be influencing, if not driving, the atmospheric chemistry, dynamics and mass-loss on all three planets. Future observations at longer IR wavelengths with JWST may provide insight into possible aerosol compositions on Kepler 51b and 51d, while optical observations may result in determining a potential Rayleigh scattering slope (beyond our two-point broadband analysis) indicative of a haze layer or a dusty-outflow. We also expect Kepler 51b in particular to be undergoing \replaced{significant atmospheric loss ($10^{10}-10^{11}$~g/s)}{atmospheric mass loss on orders of $10^{10}-10^{11}$~g/s ($10^{-2}-10^{-1}$ $M_{\oplus}$/Gyr)}, which we might hope to observe. At a distance of 800 parsecs, absorption by the interstellar medium prohibits direct Ly$\alpha$ observations of neutral hydrogen outflows, but it might be possible to observe ongoing escape with the metastable helium \replaced{line}{triplet centered} at 10833 angstroms \citep{seager.helium,helium.model,spake.et.al.2018}. \added{As noted by \citet{oklopvic.2019}, G-type stars are not expected to be promising candidates for metastable helium detections as the higher EUV flux balances the recombination and photoionziation of helium in this state. In fact, there have already been several non-detections of planets orbiting a G-type star: Wasp-12b \citep{kreidberg.and.oklopcic.2019} and HD 209458b \citep{nortmann.et.al.2018}. However, considering Kepler 51's younger age and thus higher EUV flux and activity, Kepler 51b in particular may still provide an interesting target.} Regardless, with their large scale heights and young ages, the Kepler 51 planets will continue to be unique and interesting targets for future studies.

\acknowledgments

We are extremely grateful to Tricia Royle and the staff at the Space Telescope Science Institute for their heroic efforts in scheduling these \HST observations. We thank Catherine Huitson discussing WFC3 instrumental systematics, Yifan Zhou and Daniel Apai for explaining their WFC3 charge trap model, Matteo Broggi and Eve Lee for their helpful Kepler 51 discussions, and Lile Wang and Fei Dai for sharing their dusty outflow manuscript before publication. We acknowledge both the MIT Exoplanet Tea and the CU Boulder Exoplanet Coffee for helpful discussions that motivated and improved this work. We thank Sarah Millholland for sharing her manuscript on obliquity tides before publications.

We acknowledge the anonymous reviewer whose suggestions and attention to detail both improved and strengthened this work. 

The work was based on observations made with the NASA/ESA Hubble Space Telescope, obtained from the data archive at the Space Telescope Science Institute. STScI is operated by the Association of Universities for Research in Astronomy, Inc. under NASA contract NAS 5-26555. Support for this work was provided by NASA through grant number HST-GO-14218.010-A from the Space Telescope Science Institute, which is operated by AURA, Inc., under NASA contract NAS 5-26555. J.M.D. acknowledges that the research leading to these results has received funding from the European Research Council (ERC) under the European Union's Horizon 2020 research and innovation programme (grant agreement no. 679633; Exo-Atmos). J.M.D acknowledges support by the Amsterdam Academic Alliance (AAA) Program. This material is based upon work supported by the National Science Foundation (NSF) under Grant No. AST-1413663. Support for
this work was provided by NASA through an award issued by JPL/Caltech (P90092). Work by K.M. was performed under contract with the California Institute of Technology (Caltech)/Jet Propulsion Laboratory (JPL) funded by NASA through the Sagan Fellowship Program executed by the NASA Exoplanet Science Institute.

%

\vspace{5mm}
\facilities{HST(WFC3), Kepler}


\software{astropy \citep{astropy.2013, astropy.2018},  
         photutils \citep{photutils},
         emcee \citep{mcmc.paper},
         batman \citep{batman.paper},
         isochrones \citep{morton.2015},
         ldtk \citep{ldtk.2015}
          }

\bibliography{useful_papers}


\end{document}